\documentclass[journal]{IEEEtran}
\ifCLASSINFOpdf
\else
\fi

\usepackage{graphicx}
\usepackage[T1]{fontenc}

\usepackage{algorithm}
\usepackage{lineno,hyperref,color}
\usepackage{epstopdf}

\usepackage{algpseudocode}
\usepackage{amsmath}
\usepackage{amsfonts}
\usepackage{amssymb}
\usepackage{txfonts}
\usepackage{dsfont}
\usepackage{tabu}
\usepackage{hhline}
\usepackage{color}
\usepackage{blindtext}
\usepackage{multicol}

\usepackage{xcolor}
\usepackage{wasysym}
\usepackage{caption}
\usepackage{subcaption}

\usepackage{kotex}

\definecolor{mycolor}{RGB}{0, 110, 0}
\newcommand{\jskim}{\textcolor{black}}

\usepackage{enumitem}

\def\tcb{\textcolor{black}}

\normalsize
\include{MacroFile1}
\begin{document}


\title{\jskim{\textsf{FiFo}: Fishbone Forwarding in Massive IoT Networks}}

\author{Hayoung~Seong,~\IEEEmembership{Student~Member,~IEEE,}
        Junseon~Kim,~\IEEEmembership{Student~Member,~IEEE,}
        Won-Yong~Shin,~\IEEEmembership{Senior~Member,~IEEE,}
        and~Howon~Lee,~\IEEEmembership{Senior~Member,~IEEE.}
\thanks{This research was supported in part by the National Research Foundation of Korea (NRF) grant funded by the Korea government (MSIT) (No. 2022R1A2C1010602) and in part by the National Research Foundation of Korea (NRF) grant funded by the Korea government (MSIT) (No. 2021R1A2C3004345). \it{(Corresponding authors: W.-Y. Shin; H. Lee.)}}
\thanks{H. Seong is with Department of Information Security Engineering, University of Science and Technology (UST), Daejeon 34113, South Korea (e-mail: shy2028@ust.ac.kr).}
\thanks{J. Kim is with the School of Electrical and Computer Engineering, Ulsan National Institute of Science and Technology (UNIST), Ulsan 44919, South Korea (e-mail: jskim@unist.ac.kr).} 

\thanks{W.-Y. Shin is with the School of Mathematics and Computing (Computational Science and Engineering), Yonsei University, Seoul 03722, Republic of Korea, and also with Pohang University of Science and Technology (Artificial Intelligence), Pohang 37673, Republic of Korea (e-mail: wy.shin@yonsei.ac.kr).}
\thanks{H. Lee is with School of Electronic and Electrical Engineering and IITC, Hankyong National University, Anseong, Gyeonggi 17579, South Korea (e-mail: hwlee@hknu.ac.kr).}
\thanks{{H. Seong and J. Kim are contributed equally.}}
}

\markboth
{Submitted to IEEE Internet of Things Journal}
{Submitted to IEEE Internet of Things Journal}

\maketitle
 \begin{abstract}
Massive Internet of Things (IoT) networks have a wide range of applications, including but not limited to the rapid delivery of emergency and disaster messages. Although various benchmark algorithms have been developed to date for message delivery in such applications, they pose several practical challenges such as insufficient network coverage and/or highly redundant transmissions to expand the coverage area, resulting in considerable energy consumption for each IoT device. To overcome this problem, we first characterize a new performance metric, {\em forwarding efficiency}, which is defined as the ratio of the coverage probability to the average number of transmissions per device, to evaluate the data dissemination performance more appropriately. Then, we propose a novel and effective forwarding method, \textsf{\underline{fi}shbone \underline{fo}rwarding (FiFo)}, which aims to improve the forwarding efficiency with acceptable computational complexity. Our \textsf{FiFo} method completes two tasks: 1) it clusters devices based on the unweighted pair group method with the arithmetic average; and 2) it creates the main axis and sub axes of each cluster using both the expectation-maximization algorithm for the Gaussian mixture model and principal component analysis. We demonstrate the superiority of \textsf{FiFo} by using a real-world dataset. Through intensive and comprehensive simulations, we show that the proposed \textsf{FiFo} method outperforms benchmark algorithms in terms of the forwarding efficiency.

\end{abstract}


\section{Introduction}
\label{sec_introduction}

\IEEEPARstart{W}{ith} the rapid increase in a massive number of devices, the Internet of Things (IoT) is envisioned to enable many types of communication services and applications such as autonomous driving, factory automation, and flying drone networking~\cite{MassiveIoT01, MassiveIoT02, Liu_TII}. A massive IoT network connects numerous physical devices and provides reliable wireless connectivity and autonomous management between these IoT devices. In this respect, massive IoT networks are expected to play a key role in innovating traditional industries and bringing significant benefits to society in the future. 


To support reliable wireless connectivity between things without the intervention of human and central controllers (e.g., an IoT gateway) in IoT use cases, such as disaster alarms, military applications, and mobile advertisements, \tcb{device-to-device (D2D) communications have received considerable attention. 
When a D2D technology is exploited in a massive IoT network, the behavior of this network is similar to that of a typical mobile ad hoc network.}
Accordingly, rapid and efficient data dissemination (or equivalently, message forwarding) is one of the most important and fundamental research topics in the field of massive IoT networks.

\begin{figure}[!t]
    \centering
    \includegraphics[width=0.95\columnwidth]{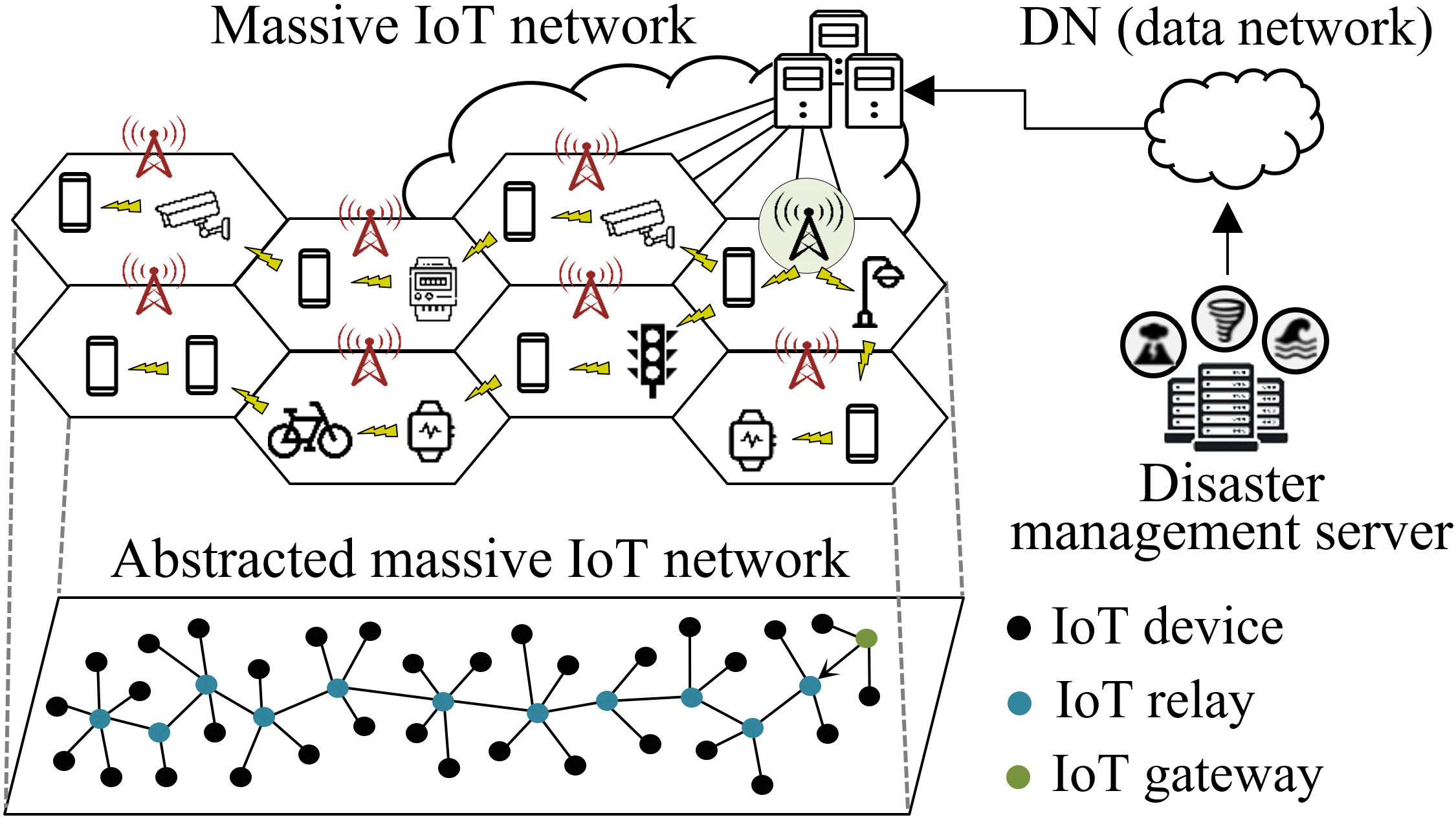}
    \caption{\jskim{A service scenario for message dissemination based on D2D communications between IoT devices with a built-in chipset capable of cellular communications in massive IoT networks.}}
    \label{fig:intro}
    \vspace{-0.4cm}
\end{figure}

Data dissemination has been designed primarily to maximize coverage probability and/or minimize duplicate transmissions~\cite{D2D1}. For example, in emergency and disaster alarm services, alarm messages must be delivered to all neighbors as soon as possible~\cite{WHAT5G}. In particular, when IoT gateways are not working temporarily owing to disasters, sudden faults, or data traffic concentrations, messages should be delivered quickly using other communication strategies such as wireless D2D communications~\cite{D2D1, D2D2}. 
In other words, D2D communication could be a good option for reliable and low-latency transmissions in such urgent situations. \tcb{Fig.~\ref{fig:intro} describes a service scenario for message dissemination based on D2D communications between IoT devices with a built-in chipset capable of cellular communications, such as mobile devices, wearable devices, and the city infrastructure, including CCTVs, streetlights, traffic lights, utility metering, and bicycles. Our scenario is valid especially when IoT gateways, such as base stations, do not work properly owing to a natural disaster. Upon detecting an urgent situation, a disaster management server transmits an emergency message to a cellular network.
The cellular network forwards messages to IoT devices within the coverage of a normal base station. Subsequently, through consecutive broadcasts of the devices, the message could be delivered to all the IoT devices that are disconnected from the cellular network.}


In disaster and emergency situations, it is crucial for IoT devices over a given network to reliably receive as much data as possible. To this end, it is necessary to design data dissemination approaches to maximize coverage probability, which is defined as the message reception ratio in the underlying network. The best strategy for coverage probability maximization is to unconditionally forward messages received from neighboring devices, regardless of the number of message transmissions. For example, in epidemic routing~\cite{epidemic}, devices broadcast messages received from neighboring devices until the entire network area is covered. However, such excessive broadcasting utilizes a large amount of transmitting power owing to the numerous message transmissions and receptions, thus resulting in a severe waste of batteries in IoT devices. Consequently, the application of data dissemination between low-power IoT devices is significantly limited.

\begin{figure}[!t]
     \centering
     \begin{subfigure}[t]{.23\textwidth}
         \centering
         \includegraphics[width=\textwidth]{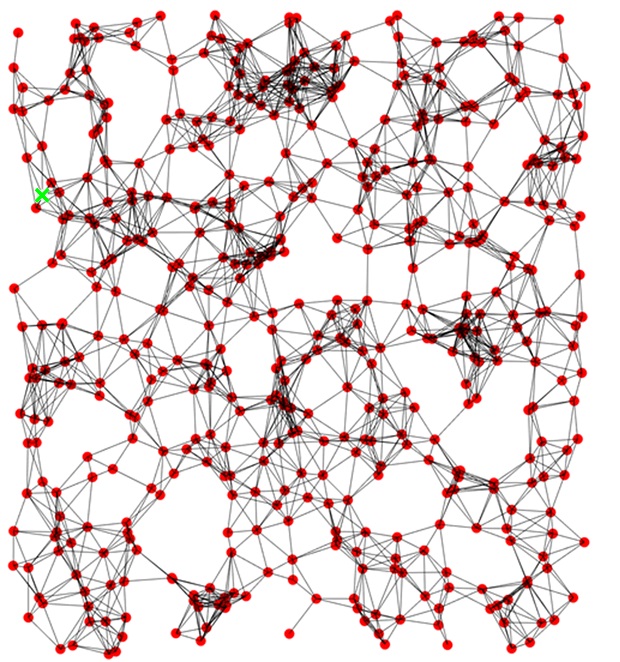}
         \caption{Epidemic routing}
         \label{DMR2A}
     \end{subfigure}
     \hspace{1mm}
     \begin{subfigure}[t]{.23\textwidth}
         \centering
         \includegraphics[width=\textwidth]{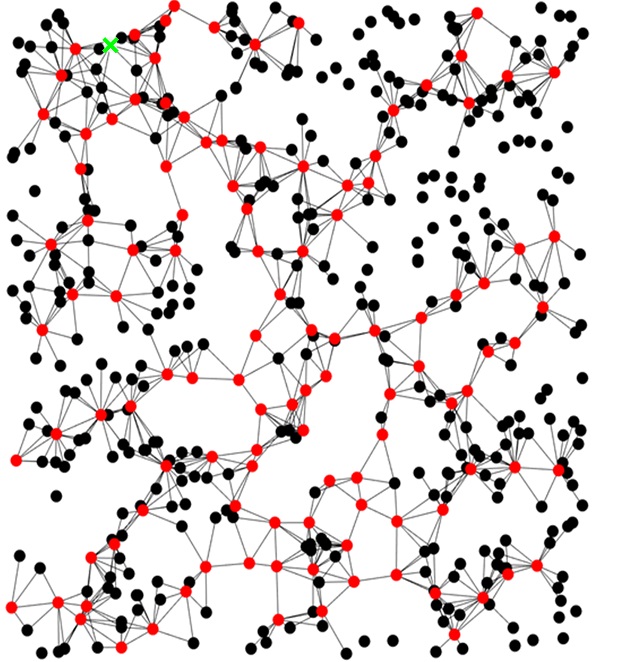}
         \caption{Algorithm with a high forwarding efficiency}
         \label{DMR2B}
     \end{subfigure}
        \caption{Data dissemination maps where the \tcb{IoT} devices with and without participation in message forwarding are marked with a red circle (`\textcolor{red}{$\bullet$}') and a black circle (`\textcolor{black}{$\bullet$}'), respectively. Here, a source device is marked with a green cross~(`\textcolor{green}{$\times$}').}
        \label{DMR2}
        \vspace{-0.4cm}
\end{figure}

To overcome this problem, our study characterizes a new performance metric, {\em forwarding efficiency}, which is defined as the ratio of coverage probability to the average number of transmissions per device. \tcb{This new metric enables us to more appropriately evaluate the performance of data dissemination in massive IoT networks.} For example, let us consider two different data dissemination algorithms. Fig.~\ref{DMR2} illustrates the data dissemination map added when epidemic routing~\cite{epidemic} and another algorithm leading to a high forwarding efficiency are employed. It can be observed that as shown in Fig.~\ref{DMR2}(a), most of the devices participate in message forwarding when epidemic routing is used, which is in sharp contrast to the case where an algorithm with high forwarding efficiency is used because only parts of the devices are active for transmissions (see Fig.~\ref{DMR2}(b)). Our aim is to design a high-forwarding-efficiency algorithm suitable for low-power IoT use cases while guaranteeing low computational complexity. \tcb{The forwarding efficiency jointly considers the network-wide energy consumption and coverage probability so that it can be one of the most important performance metrics in massive IoT networks. In particular, because energy consumption is closely related to the number of transmissions and receptions, reducing message duplicates can contribute to an increase in network-wide energy efficiency.}

In this paper, we propose a novel data dissemination framework, \textsf{\underline{fi}shbone \underline{fo}rwarding (FiFo)}, which improves forwarding efficiency without causing the problem of computational complexity issues for massive IoT networks. \textsf{FiFo} utilizes the location information of users obtained from cellular systems. Based on location information, \textsf{FiFo} performs the following two steps. First, devices in the underlying network are clustered based on the unweighted pair group method with arithmetic average (UPGMA)~\cite{UPGMA}. Second, a fishbone discovery strategy is employed. This strategy is divided into two phases: 1) creation of the main axis and 2) creation of the sub axes. The main axis and sub axes of each cluster correspond to the spine and thorns of the fishbone, respectively, and are created by using both the expectation-maximization (EM) algorithm for the Gaussian mixture model (GMM) and principal component analysis (PCA).

To numerically validate the performance of the proposed \textsf{FiFo} method over benchmark data dissemination methods, we perform comprehensive numerical evaluations using a real-world dataset. The experimental results demonstrate that our \textsf{FiFo} method significantly outperforms all benchmark algorithms with respect to forwarding efficiency. Moreover, motivated by the observation that the transmission range covered by IoT devices may be limited owing to potential coverage holes, we consider the case where an additional relay is deployed as a replenishment step built upon the \textsf{FiFo} method, which enables us to forward messages to devices within each coverage hole. From the replenishment step, we show that when our \textsf{FiFo} method is used, the enhancement of the coverage probability is noticeable by adding only one additional relay while still achieving the highest forwarding efficiency among all methods. Finally, we empirically show that our \textsf{FiFo} method exhibits acceptable runtime complexity compared with other benchmark methods.

\jskim{
Our contributions are summarized as follows:
\begin{itemize}
    \item We propose a holistic topology-aware data dissemination framework for massive IoT networks that effectively acquire urgent and crucial information such as device locations and disaster information in urgent situations.
    \item 
    We develop a novel analytical technique that discovers forwarding paths in the shape of a fishbone based on analyzing the network-wide device distribution. The proposed fishbone-shaped forwarding enables us to perform efficient and rapid data dissemination by reducing the number of duplicate relays and their overlapping coverage regions.
    \item Through intensive performance evaluation, we demonstrate that \textsf{FiFo} outperforms the existing algorithms such as epidemic routing~\cite{epidemic}, broadcast incremental power (BIP)~\cite{BIP}, probabilistic flooding (PF)~\cite{BroadcastStorm,BulkDataDissemination}, and neighbor-based probabilistic broadcast (NPB)~\cite{NPB} with respect to forwarding efficiency. Furthermore, we discuss the performance behavior of several \textsf{FiFo} variants to compare the performance of the proposed \textsf{FiFo} with various topology-aware broadcasting strategies exploiting the status of neighboring nodes.
\end{itemize}
}

The remainder of this paper is organized as follows. 
\jskim{In Section~\ref{sec_motivation}, we present key requirements motivated by efficient and rapid data dissemination in massive IoT networks. We then provide an overview of the \textsf{FiFo} data dissemination framework in Section~\ref{sec_overview} and explain the technical details in Section~\ref{sec_fishbone}.} In Section~\ref{sec_dataset}, we describe a real-world dataset. Comprehensive numerical results are presented in Section~\ref{sec_results}. In Section~\ref{sec_model}, we present prior studies related to our work. Finally, we provide a summary and concluding remarks in section ~\ref{sec_conclusion}.

\section{Motivation}
\label{sec_motivation} 

\jskim{In this section, we address the motivation of our study by identifying two key requirements for rapid and efficient data dissemination in massive IoT networks. Based on these requirements, we design an energy-efficient fishbone-shaped forwarding strategy, called \textsf{FiFo}, as explained in Sections ~\ref{sec_overview} and~\ref{sec_fishbone}. These two essential requirements are described as follows.}

\vspace{-0.5cm}

\jskim{\subsection {Full Topology-Aware Data Dissemination}
To deliver emergency messages to all devices in a timely manner, it is crucial to judiciously determine the relay devices. 
    A simple method is epidemic routing~\cite{epidemic} in which all devices rebroadcast messages received from neighboring devices.
    %
    However, this method yields severe energy waste because each device must perform many unnecessary transmissions and receptions.
    Owing to the heavy traffic generated by the concurrent transmissions of adjacent devices over the same radio channel in epidemic routing, many collisions occur.
    %
    %
    Packet collisions can lead to undesired retransmissions that exacerbate the energy inefficiency problem, and worse, messages may not be delivered successfully. 
    To alleviate this problem, in a recent study, NPB~\cite{NPB} introduced a method that selects relay devices in a stochastic manner while considering the number of devices not receiving the message.
    %
    However, it is not easy to forward a message efficiently to widely distributed devices because, in selecting a relay device, only the surrounding network topology is utilized rather than the entire topology. 
    Therefore, we herein develop a novel data dissemination algorithm that takes advantage of the topological information of the entire network, which enables messages to be delivered to almost all devices with high energy efficiency. }

\jskim{\subsection {Efficient Forwarding Information Acquisition}
Network-wide topological information can be exploited in massive IoT networks to efficiently find forwarding paths.
    %
    %
    Recent work on wireless sensor networks (WSNs) aims to devise an energy-efficient forwarding algorithm~\cite{cent_wsn}. To this end, a source with a centralized controller collects information regarding the energy level of each node, which is embedded into the sensor data to be exchanged.
    However, this method requires considerable time to gather information from all devices as the network grows in scale. Furthermore, in the case of variations in the network topology, the collected information may be useless. 
    To efficiently and quickly obtain forwarding information, we design a cellular-aided data dissemination framework that utilizes device locations periodically tracked for mobility management in cellular networks~\cite{sdn_nfv, oran}. }

\section{\textsf{FiFo} Overview }
\label{sec_overview}

\begin{figure}[!t]
     \centering
     \begin{subfigure}[b]{1.0\columnwidth}
         \centering
         \includegraphics[width=0.95\columnwidth]{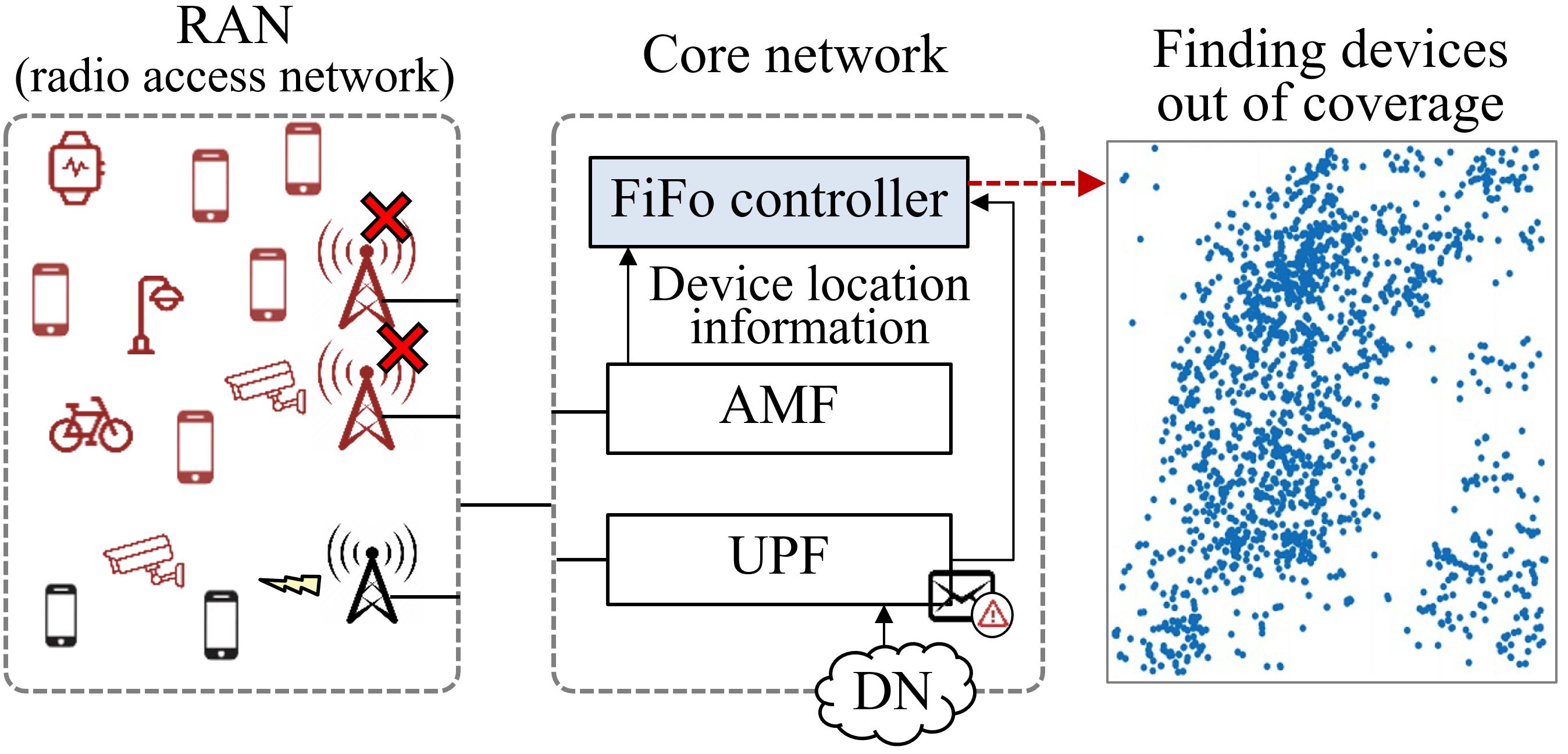}
         \caption{\jskim{IoT device discovery within the cellular coverage upon detecting the disaster at the FiFo controller}}
         \vspace{0.3cm}
     \end{subfigure}
     \begin{subfigure}[b]{1.0\columnwidth}
         \centering
         \includegraphics[width=0.95\columnwidth]{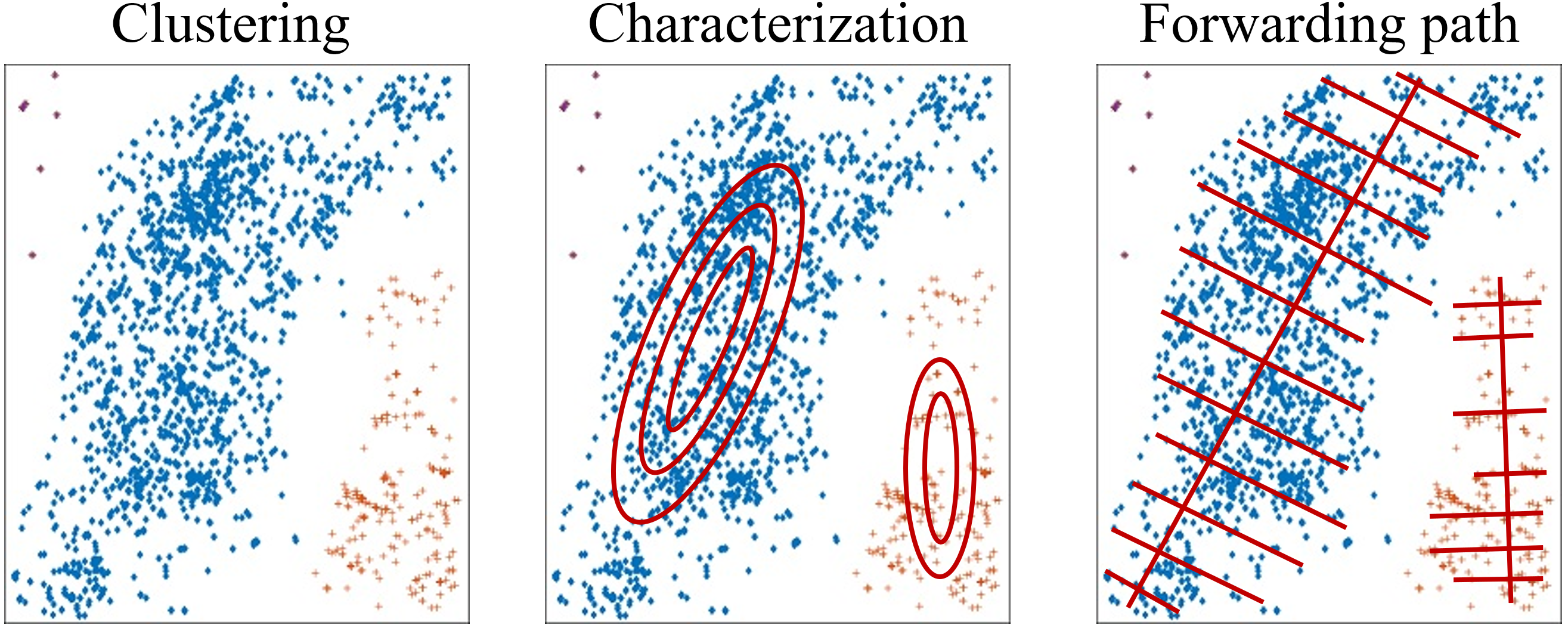}
         \caption{\jskim{Fishbone-shaped forwarding path analysis based on the device distribution}}
         \vspace{0.3cm}
     \end{subfigure}
     \begin{subfigure}[b]{1.0\columnwidth}
         \centering
         \includegraphics[width=0.95\columnwidth]{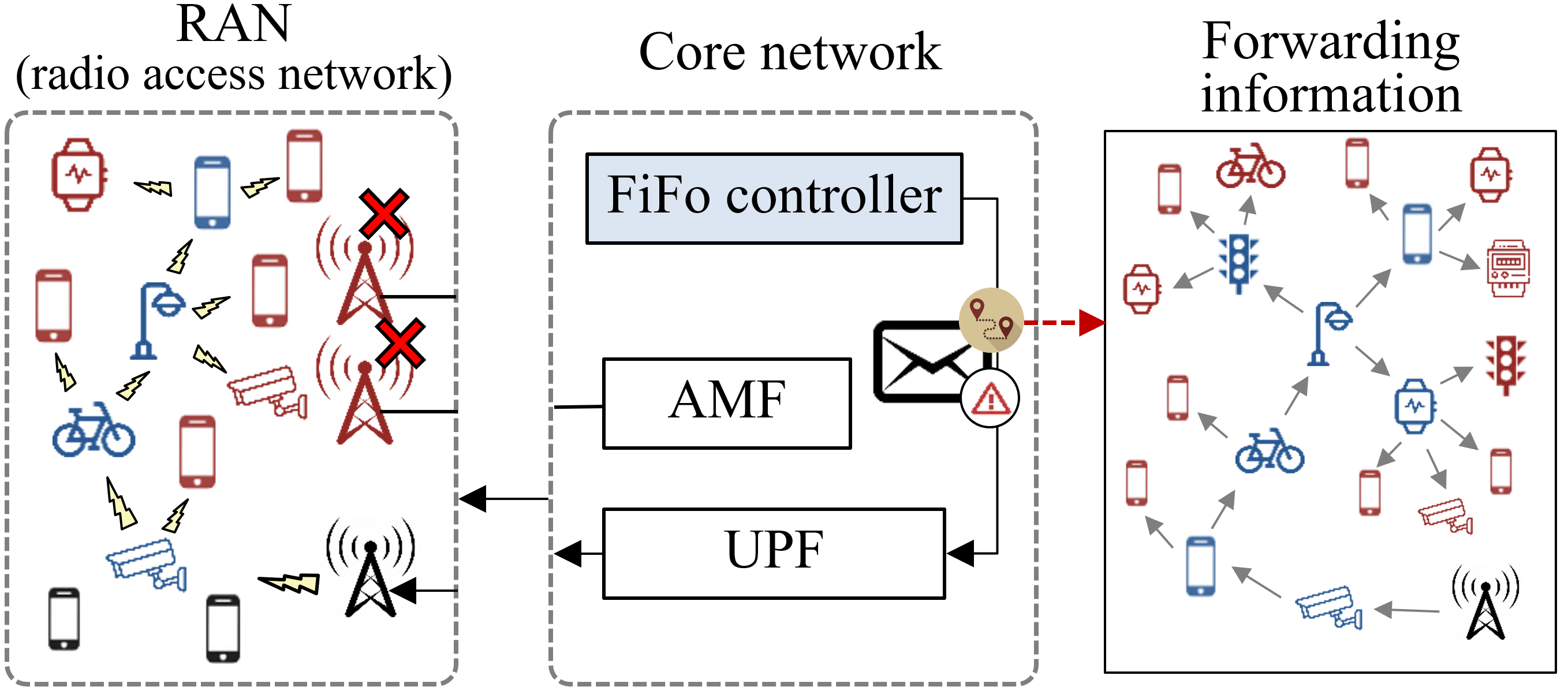}
         \caption{\jskim{Data dissemination based on IoT relays selected to forward data along the fishbone-shaped forwarding path}}
    \end{subfigure}
    \caption{\jskim{An overview of the proposed \textsf{FiFo} framework.}}
    \label{fig:overview}
    \vspace{-0.4cm}
\end{figure}

\jskim{
\textsf{FiFo} is a new cellular-aided data dissemination framework for rapidly and efficiently disseminating a message to all the devices in massive IoT networks, as shown in Fig.~\ref{fig:overview}.
\textsf{FiFo} makes use of the location information of devices, which is periodically collected by the key entity of cellular systems, such as the mobility management entity (MME) in LTE and the access and mobility management function (AMF) in 5G~\cite{lmf}. Based on the location information, \textsf{FiFo} performs two key functions for fast and efficient forwarding in massive IoT networks: 1) forwarding path analysis and 2) data dissemination to spread a message to all devices based on the analyzed forwarding path.
}

\jskim{
When the \textsf{FiFo} controller receives an emergency message from the disaster management server, it first searches for devices that no longer receive a cellular service because of the failure of base stations. This procedure can be performed simply by identifying devices within the coverage of faulty base stations. Then, using the location information of the devices obtained from the AMF, we analyze the established forwarding paths in the shape of a fishbone, which enable us to maximize the forwarding coverage with high energy efficiency. Based on the analyzed forwarding path, the controller judiciously selects a set of relay devices and delivers an emergency message with information about the selected relays via the user plane function (UPF). A device chosen as the relay rebroadcasts the received message to neighboring devices.
}

\jskim{
When analyzing the fishbone-shaped forwarding path, the controller first groups the devices (i.e., clustering) in terms of their geographical positions. The reason is that, while it is better to extend one route between adjacent devices, for users who are far away, it can be more efficient to create additional routes. After clustering, to generate a forwarding skeleton according to the distribution of each cluster, the controller finds the main axis (i.e., a spine) and sub axes (i.e., thorns) by analyzing the dispersion degree of the distribution in the two-dimensional domain. This analysis is performed using PCA which utilizes the distribution characterized by the EM algorithm with the GMM. Finally, the controller selects relay devices to forward messages along the forwarding skeleton.
}

\jskim{
Once a set of relays is determined, the \textsf{FiFo} controller begins to propagate the emergency message, including the relay information, through the normal IoT gateway.
When a device in the coverage area of the IoT gateway receives the message, the device checks whether it has been selected as a relay. If selected, then the relay device transmits the message to the neighboring devices by excluding itself from the relay-device information. If not selected, then the device does not send a message.
}

\section{\textsf{FiFo} Framework} \label{sec_fishbone} 

\jskim{
We describe the fishbone-shaped forwarding path analysis and data dissemination algorithm of \textsf{FiFo} in detail.
}


\subsection{Forwarding Path Analysis}

\tcb{We present our \textsf{FiFo} method in massive IoT networks, which is composed of three phases: 1) device clustering, 2) creation of a main axis, and 3) creation of sub axes.}
%

\subsubsection{Phase 1: Device Clustering} \label{sec_clustering}

In Phase 1 of the \textsf{FiFo} method, we perform device clustering based on the UPGMA~\cite{UPGMA}, which is a simple agglomerative hierarchical clustering strategy built upon the matrix of pairwise distances between users in the underlying network. In our study, to build a matrix of pairwise distances using the Twitter dataset, we adopt the Mahalanobis distance $D_{M}({{\bf p}_i},{{\bf p}_j})$, which is a generalized squared interpoint distance between two positions ${\bf p}_i$ and ${\bf p}_j$.
where ${\bf p}_i\in \varmathbb{R}^2$ denotes the geographical position of user $i$ in the Cartesian coordinate system. Then,
$D_{M}({{\bf p}_i},{{\bf p}_j})$ is represented as
\begin{align}\label{Dmh}
D_{M}({{\bf p}_i},{{\bf p}_j})=\sqrt{\left({{\bf p}_i}-{{\bf p}_j}\right) {\bf{\Sigma}}^{-1} \left({{\bf p}_i}-{{\bf p}_j}\right)^T},
\end{align}
where $\bf{\Sigma}$ is the sample covariance matrix of user locations over the network and $T$ is the transpose of a vector. 
Using a dissimilarity matrix constructed from equation ($\ref{Dmh}$), the UPGMA enables us to build a rooted dendrogram. In each step of the UPGMA, the nearest two clusters are combined into a high-level cluster. The overall procedure of the UPGMA algorithm is as follows.
\begin{enumerate}[label=(\alph*)]
    \item Find a pair $(i,j)$ of two elements with the smallest distance in the dissimilarity matrix;
    \item Construct a new high-level cluster $k$ comprising the two elements $i$ and $j$;
    \item Update the distance between the new cluster $k$ and other elements;
    \item Remove the elements $i$ and $j$ and then add the new cluster $k$ into the tree;
    \item If $i$ and $j$ were the last 2 clusters, then this procedure is terminated. Otherwise, go back to the first step.
\end{enumerate}

\begin{figure}[!t]
     \centering
     \begin{subfigure}[b]{.2\textwidth}
         \centering
         \includegraphics[width=4cm]{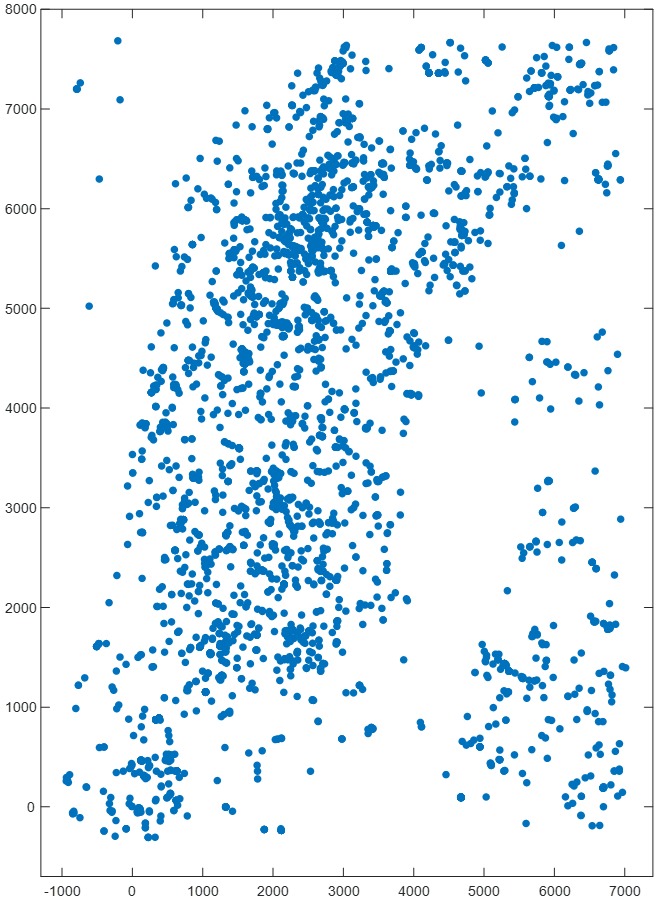}
         \caption{Before clustering}
         \label{clusteringA}
     \end{subfigure}
     \hspace{5mm}
     \begin{subfigure}[b]{.2\textwidth}
         \centering
         \includegraphics[width=4cm]{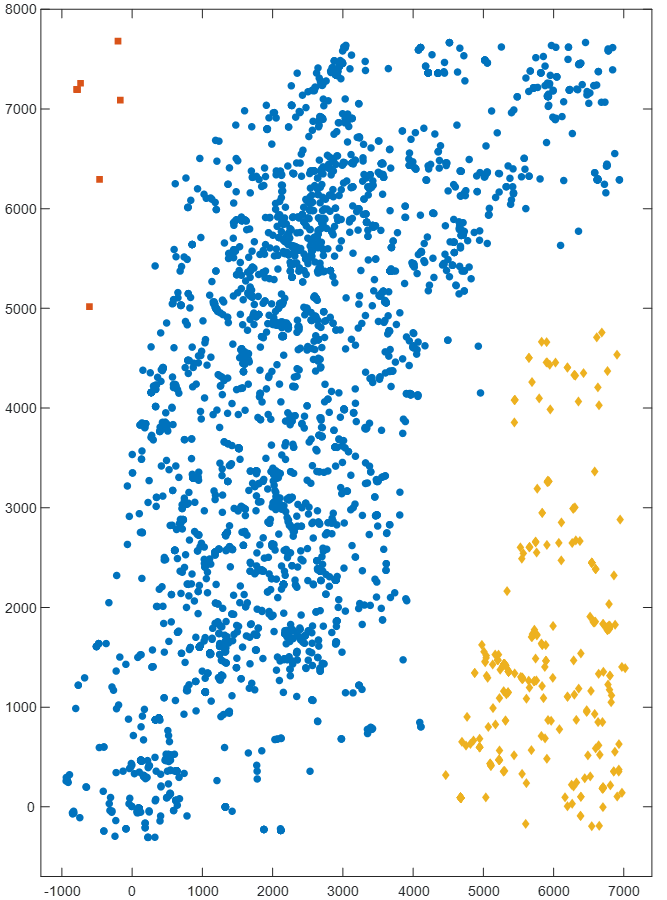}
         \caption{After clustering}
         \label{clusteringB}
     \end{subfigure}
        \caption{\tcb{IoT-device} clustering based on the UPGMA over a down-town region in New York City, \tcb{having an area of 8$km^2$, where the x- and y-coordinates represent the longitude and latitude, respectively.}} 
        \label{clustering}
\end{figure}

Fig.~\ref{clustering} shows the clustering result using the UPGMA algorithm, \tcb{where the real-world dataset obtained via Twitter corresponding to the downtown region in New York City with an area of 8$km^2$ is used (see Section~\ref{sec_dataset} for more details). In the figure, the x- and y-coordinates represent the longitude and latitude, respectively, and for the numerical evaluation, the map in the figure is resized to $8,700\times 8,700$. Figs.~\ref{clustering}(a) and~\ref{clustering}(b) illustrate the device distributions before and after clustering, respectively, when the number of clusters is assumed to be three.} The entire region is divided into three clusters, \tcb{including the upper left (smallest cluster), middle (largest cluster), and lower right,} each of which is marked by its unique color.

\subsubsection{Phases 2 and 3: Discovering a Fishbone}
\label{sec_axes}

\begin{figure}[!t]
     \centering
     \begin{subfigure}[t]{.24\textwidth}
         \centering
         \includegraphics[width=\textwidth]{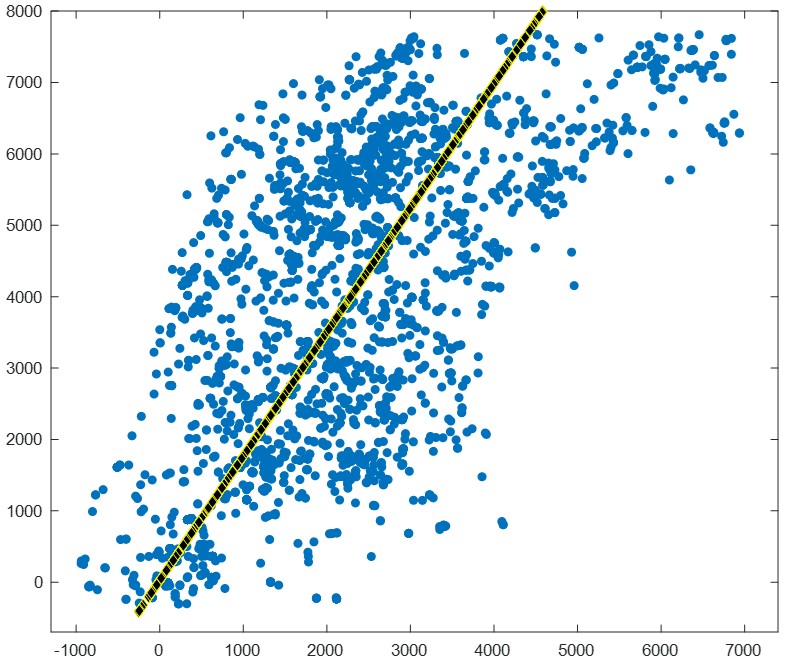}
         \caption{Creation of the main axis}
         \label{makingaxisA}
     \end{subfigure}
     \hspace{-1mm}
     \begin{subfigure}[t]{.24\textwidth}
         \centering
         \includegraphics[width=\textwidth]{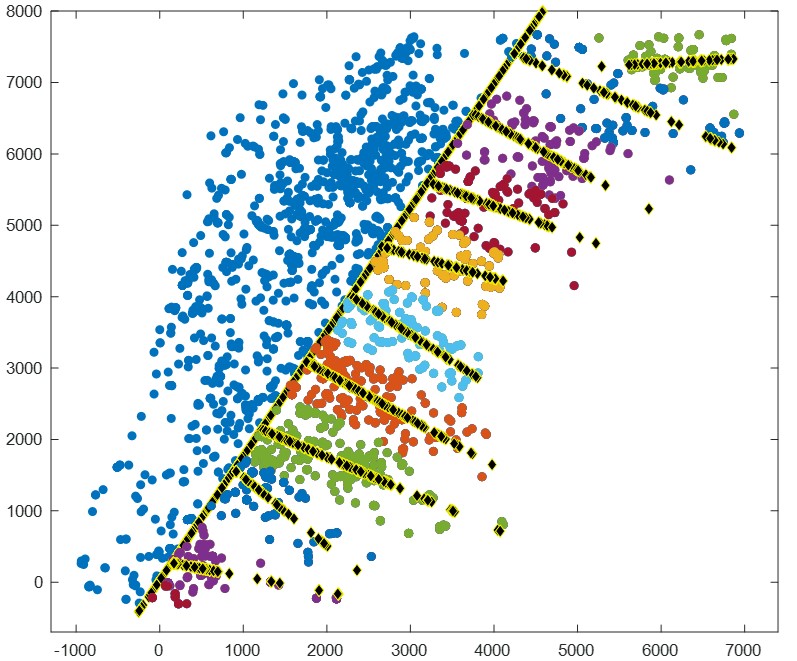}
         \caption{Creation of sub axes}
         \label{makingaxisB}
     \end{subfigure}

        \caption{Discovery procedure of the main axis and sub axes for the largest cluster shown in Fig.~\ref{clustering}.}
        \label{makingaxis}
\end{figure}

In this subsection, we present our fishbone discovery strategy, which is partitioned into the following two phases: $^{1)}$creation of the main axis and $^{2)}$creation of sub axes. Note that the main axis and sub axes correspond to the spines and thorns of the fishbone, respectively. Fig.~\ref{makingaxis} shows an illustrative example of the overall procedure of fishbone discovery for the largest cluster shown in Fig.~\ref{clustering}. As depicted in Fig.~\ref{makingaxis}(a), the main axis of the largest cluster is represented as the major axis of the ellipse. After discovering the main axis, we also find sub axes that are perpendicular to the main axis, as shown in Fig.~\ref{makingaxis}(b). \tcb{Here, in Fig.~\ref{makingaxis}, points (corresponding to IoT devices) belonging to different clusters are marked with different colors. Furthermore, in Fig.~\ref{makingaxis}(b), IoT devices belonging to different forwarding regions partitioned according to the sub axes (i.e., devices with different subaxis groups) are marked with different colors. }

\smallskip \noindent
\textbf{Creation of a main axis: }
After clustering the devices based on the UPGMA, we aim to discover the main axis in phase 2 of the \textsf{FiFo} method. To this end, we need to find \tcb{the centroid}, covariance matrix, and eigenvector of each cluster. \tcb{Here, the centroid indicates the center point of a cluster.} If we find the eigenvector of the corresponding cluster, then all devices are projected onto the eigenvector of the target group such that the main axis is found.

In our study, we perform an EM algorithm in the context of the GMM in order to find both \tcb{the centroid} and the covariance matrix of each cluster. \tcb{The centroid and covariance matrix are calculated using the locations of the devices in each cluster, where the centroid and covariance matrix for the $k$th cluster are denoted as ${\mu}_{k}$ and ${\sigma_{k}}$, respectively.} Then, we compute both the eigenvalue and eigenvector of each cluster through PCA using the covariance matrix, \tcb{where the eigenvalue and eigenvector for the $k$th cluster are denoted as $\lambda_{k}$ and ${\bf v}_k$, respectively.} The EM algorithm consists of an E-step, which calculates the expectation value for log-likelihood parameters as an estimate of the parameters, and an M-step, which calculates parameters maximizing the expected log-likelihood parameters obtained from the E-step. We assume that the geographical positions of users are a multivariate normal distribution of a 2-dimensional random vector ${\bf X}_i\in\varmathbb{R}^2$. First, the marginal probability $P({\bf X}_i = \bf{x})$ using GMM can be expressed as
\begin{align}\label{GMM}
P({\bf X}_i = {\bf x}) = \sum_{k=1}^{K} {\bf \pi}_{k} \cdot \mathcal{N} \left({\bf X}_i = {\bf x}|~\! {\bf \mu}_{k}, {\bf \Sigma}_{k}\right),
\end{align}
where $\bf{\mu}_{k}$ and $\bf{\Sigma}_{k}$ are the mean and variance, respectively, of the $k$th Gaussian distribution $\mathcal{N(\cdot)}$, and ${\bf \pi}_k$ is the mixing coefficient. In (\ref{GMM}), conditional probability $P\left({\bf X}_i|~\! {\bf \mu}_k,{\bf \Sigma}_k\right)$ is expressed as follows:
\begin{align}\label{GMM_N}
\displaystyle{P\left({\bf X}_i|~\! {\bf \mu}_k,{\bf \Sigma}_k\right) = \frac{1}{\sqrt{2 \cdot {\bf \pi}_k|{\bf \Sigma}_k|}}\exp\left(-\frac{1}{2}{\bf \Sigma}_k^{-1}\left({\bf x}-{\bf \mu}_k\right)^{2}\right)}.
\end{align}
Here, $\sum_{k=1}^{K}{\bf \pi}_{k} = {\bf 1}$ and ${\bf 0}\leq{\bf \pi}_{k}\leq {\bf 1}$, where ${\bf 0}$ and ${\bf 1}$ represent all-zero and all-one vectors, respectively. Using (\ref{GMM}) and (\ref{GMM_N}), the log-likelihood function is given by
\begin{align}\label{GMM3_1}
\ln{ P{\left({\bf X}_i|~\!{\bf \pi}_k,{\bf \mu}_k,{\bf \Sigma}_k\right)}}=\sum_{i=1}^{N}\ln{\left\{\sum_{k=i}^{K} {\bf \pi}_{k} \cdot \mathcal{N} \left({\bf X}_i = x|~\!{\bf \mu}_{k},{\bf \Sigma}_{k}\right)\right\}}.
\end{align}

Using (\ref{GMM})--(\ref{GMM3_1}), the E-step and M-step are alternately performed as follows:
%
\begin{itemize}
    \item \textbf{E-step} (expectation): With the current ${\bf \mu}_k$ and ${\bf \Sigma}_k$, the posterior probability ${\gamma\left(z_{k}\right)}$ can be calculated by
\begin{align}\label{GMM4}
\gamma\left(z_{k}\right) 
= \frac{{\bf \pi}_{k} \cdot \mathcal{N} \left({\bf X}_i = x|~\!{\bf \mu}_{k},{\bf \Sigma}_{k}\right)}{\sum_{j=1}^{K}{\bf \pi}_{j} \cdot \mathcal{N} \left({\bf X}_i = {\bf x}|~\!{\bf \mu}_{j},{\bf \Sigma}_{j}\right)},
\end{align}
where ${\gamma\left(z_{k}\right) }$ indicates the probability that ${\bf X}_i={\bf x}$ belongs to the $k$th cluster and $z_k$ is the latent variable.
    \item \textbf{M-step} (maximization): The parameters $\hat{{\bf \mu}_{k}}$, $\hat{{\bf \Sigma}_{k}}$, and $\hat{{\bf \pi}_{k}}$ are recalculated to maximize the expected log-likelihood as follows.
\begin{align}\label{GMM5}
\hat{{\bf \mu}_{k}} & = \frac{1}{N_{k}}\sum_{i=1}^{N}\gamma\left(z_{k}\right){\bf X}_{i} \\
\hat{{\bf \Sigma}_{k}} & = \frac{1}{N_{k}}\sum_{i=1}^{N}\gamma\left(z_{k}\right)\left({\bf X}_{i}-\hat{{\bf \mu}_{k}}\right)\left({\bf X}_{i}-\hat{{\bf \mu}_{k}}\right)^{T} \\
\hat{{\bf \pi}_{k}} & = \frac{N_{k}}{N},
\end{align}
where $N_{k}$ represents $\sum_{i=1}^{N}\gamma\left(z_{k}\right)$. Then, we iterate the E-step and M-step until convergence.
\end{itemize}

After finding the \tcb{centroid} and covariance matrix of each cluster through the EM algorithm in the context of the GMM, we calculate the eigenvalue and eigenvector of each cluster using PCA, which is a widely used dimensionality reduction technique~\cite{EMPCA, EMPCA2}. PCA uses an orthogonal transformation to convert a set of devices of possibly correlated variables into a set of values of linearly uncorrelated variables, the so-called principal components. The first principal component has the largest possible variance, and each succeeding component, in turn, has the highest variance while maintaining orthogonality with the preceding components. For example, in the proposed \textsf{FiFo} method, the main axis can be determined by calculating the eigenvectors and eigenvalues of the covariance matrix ${\bf \Sigma}_k$ of the largest cluster, which includes most devices (see Fig.~\ref{makingaxis}). Then, eigenvalue ${\lambda_k}$ and eigenvector ${\bf v}_k$ can be determined according to the following equation:
\begin{align}\label{GMM7}
\left({\bf \Sigma}_k - \lambda_k \cdot {\bf I} \right){\bf v}_k=0,
\end{align}
where ${\bf I}$ denotes the identity matrix.

\smallskip \noindent
\textbf{Creation of sub axes: }
In Phase 3 of the \textsf{FiFo} method, we are interested in creating sub axes for the given main axis. To this end, we present the following three steps: (i) determination of the forwarding region according to the sub axes, (ii) selection of relay devices, and (iii) search for the rotation angle.

\begin{figure}[!t]
     \centering
     \begin{subfigure}[t]{.1126\textheight}
         \centering
         \includegraphics[width=\textwidth]{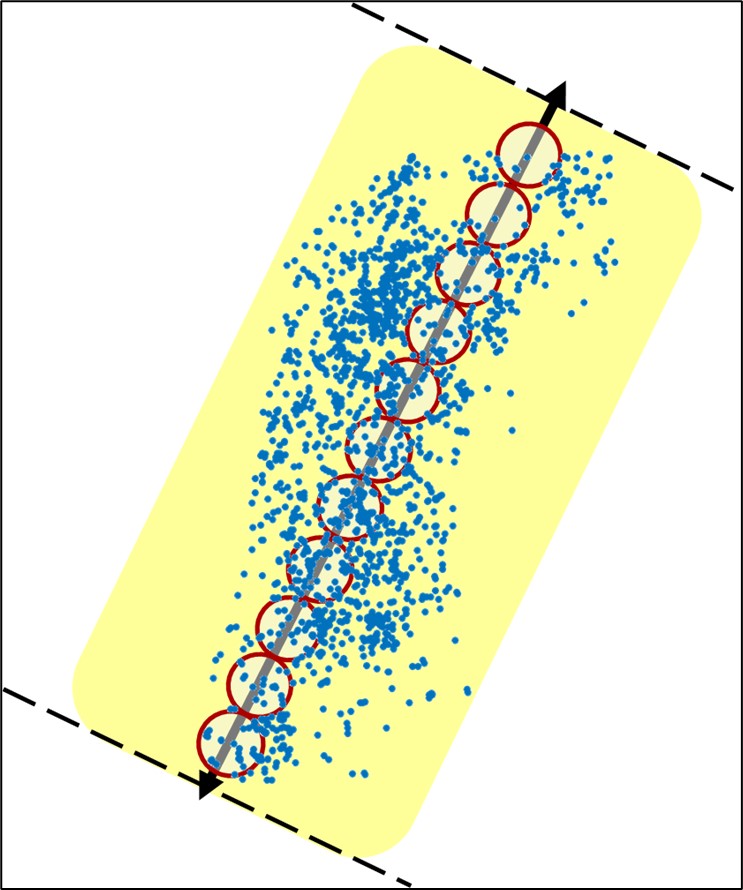}
         \caption{Main axis for the largest cluster}
         \label{makingsubaxesA}
     \end{subfigure}
     \hspace{0mm}
     \begin{subfigure}[t]{.11\textheight}
         \centering
         \includegraphics[width=\textwidth]{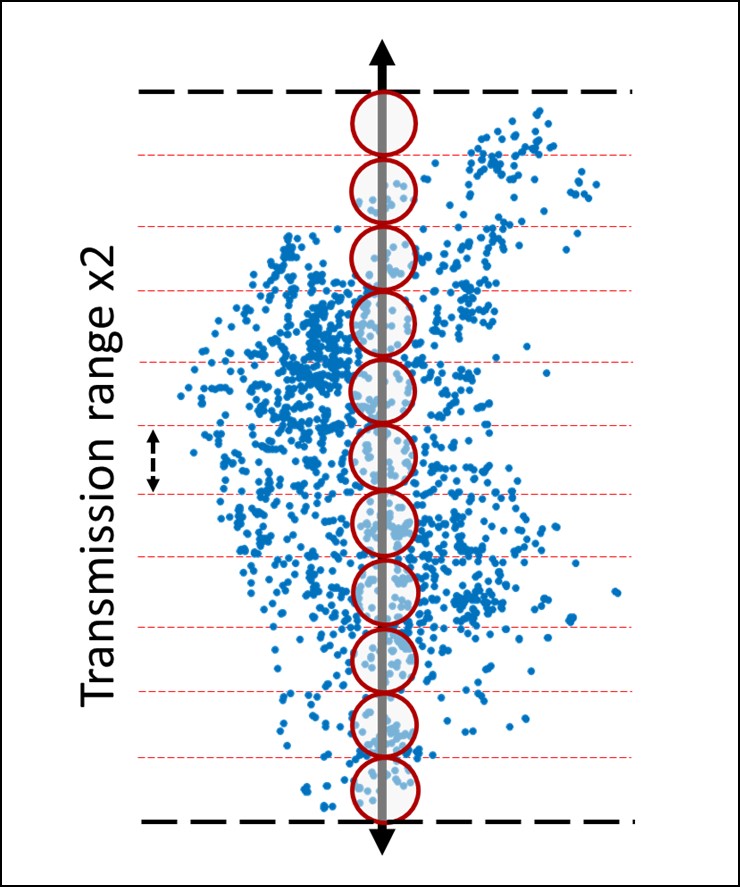}
         \caption{Rotation of the main axis by 90$^{\circ}$}
         \label{makingsubaxesB}
     \end{subfigure}
     \hspace{0mm}
     \begin{subfigure}[t]{.11\textheight}
         \centering
         \includegraphics[width=\textwidth]{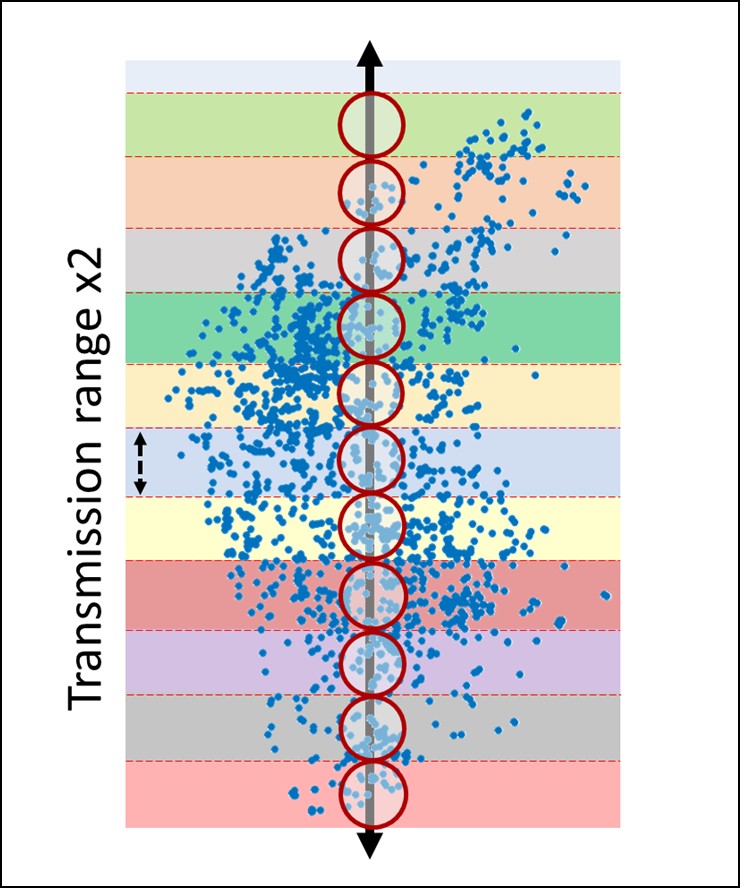}
         \caption{Partitioning the sub axes}
         \label{makingsubaxesC}
     \end{subfigure}
     
        \caption{Discovery of sub axes for the largest cluster.}
        \label{makingsubaxes}
\end{figure}

Fig.~\ref{makingsubaxes} illustrates the method used to determine the forwarding region after discovering the main axis for the largest cluster, as shown in Fig.~\ref{clustering}. To make the design tractable, we rotate the main axis such that the initial angle of the sub axes is set as $90\,^{\circ}$, as shown in Fig.~\ref{makingsubaxes}(b). It should be noted that this angle can be adaptively changed according to the device distribution of each cluster. Then, the main axis is divided such that the entire forwarding region is partitioned into disjoint sub-areas, each of which has twice the transmission range. Fig.~\ref{makingsubaxes}(c) illustrates the set of partitioned forwarding regions according to the sub axes when the angle of the sub axes is set to $90\,^{\circ}$.

\begin{figure}[!t]
     \centering
     \begin{subfigure}[b]{.17\textwidth}
         \centering
         \includegraphics[width=\textwidth]{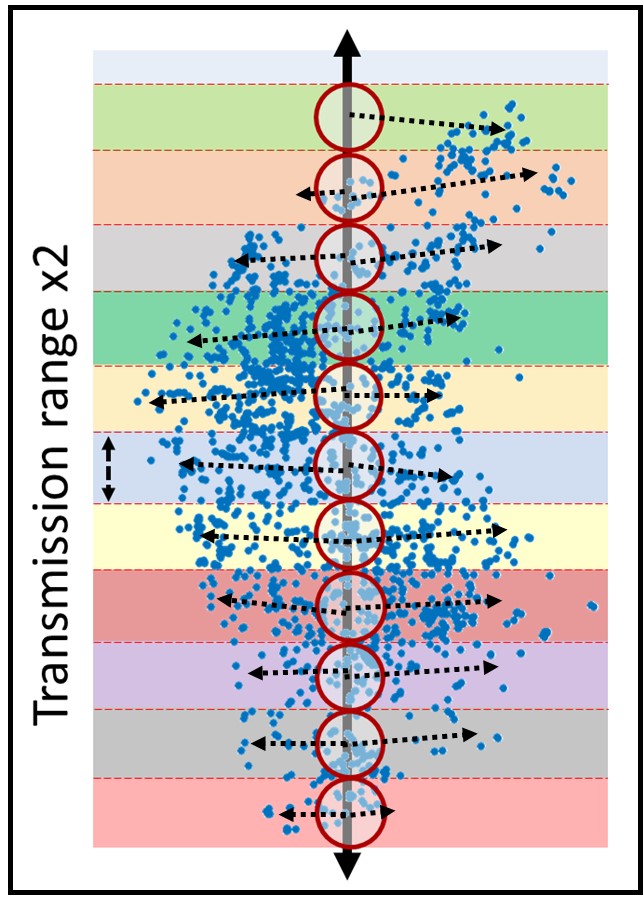}
         \caption{Partitioned sub axes}
         \label{relaynodeA}
     \end{subfigure}
     \begin{subfigure}[b]{.25\textwidth}
         \centering
         \includegraphics[width=\textwidth]{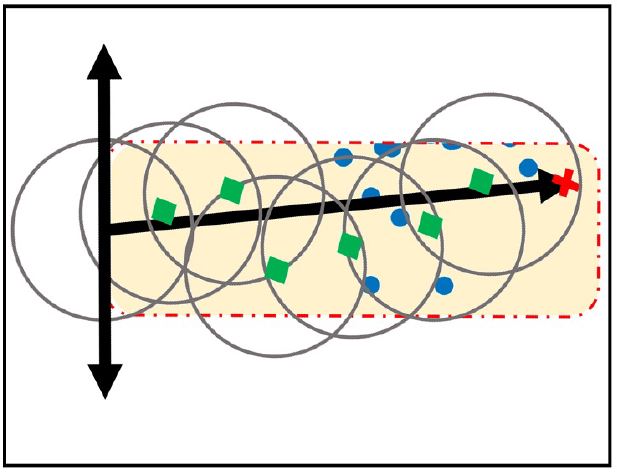}
         \caption{Relay node selection}
         \label{relaynodeB}
     \end{subfigure}

        \caption{Relay node selection in each sub axis region.}
        \label{relaynode}
\end{figure}


\begin{figure}[!t]
\centering
\includegraphics[width=7.5cm]{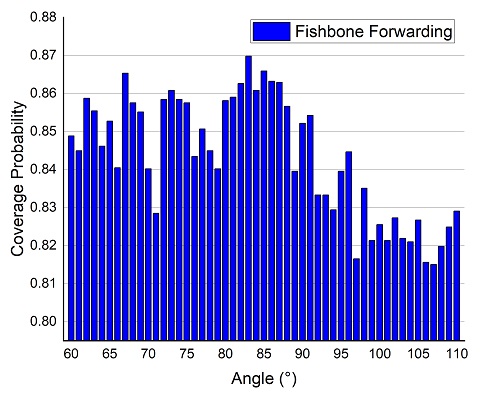}
\caption{Coverage probability according to the rotation angle of the main axis.}
\label{rotation_angle}
\end{figure}

\begin{algorithm}[!t]
\centering
\caption{\textsf{FiFo}} \label{DDA_1}
\begin{algorithmic}[1]

\item[]
\textbf{\item[\#]{Input}}
\State $\{{\bf p}_1,\cdots,{\bf p}_{|\mathcal{N}|}\}$
\textbf{\item[\#]{Output}}
\State $ MainAxis$, $ \{SubAxes(k)\}_{k=1}^{K} $ 
\textbf{\item[\#] {Clustering}}
\For { $ i = 1 : |\mathcal{N}| $ }
\For { $ j = 1 : |\mathcal{N}| $ }
\State Compute the Mahalanobis distance $D_M ({\bf p}_i,{\bf p}_j)$
\EndFor
\EndFor
\State $\{U_1,\cdots, U_K\}$ $\leftarrow$ UPGMA$(\{D_M(\cdot,\cdot)\})$
\textbf{\item[\#] {Fishbone discovery}}
\For { $ k = 1 : K $ }
\State $ ({\bf \mu}_k, {\bf \Sigma}_k) $ $\leftarrow$ GMM$(U_k) $
\State $ MainAxis$ $\leftarrow$ PCA$(\mu_k, \Sigma_k, U_k) $
\State Search for the rotation angle for creating sub axes
\State Compute $k_{sub}$, \# of sub axes in cluster $k$
    \For { $ \tilde{k} = 1 : 2k_{sub} $ }
    \State $ ({\bf \mu}_{\tilde{k}}, {\bf \Sigma}_{\tilde{k}}) $ $\leftarrow$ GMM$(U_{\tilde{k}} ) $
    \State $ SubAxes(k) $ $\leftarrow$ PCA$({\bf \mu}_{\tilde{k}}, {\bf \Sigma}_{\tilde{k}}, U_{\tilde{k}}) $    
    \EndFor
\EndFor

\end{algorithmic}
\end{algorithm}

\begin{figure*}[!t]
     \centering
     \begin{subfigure}[b]{.32\textwidth}
         \centering
         \includegraphics[width=0.75\textwidth]{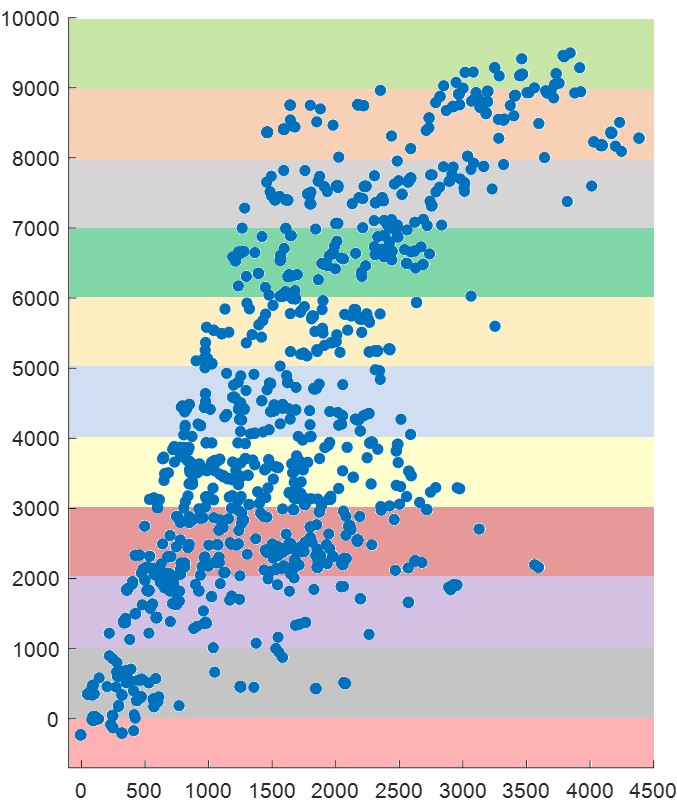}
         \caption{80$^{\circ}$}
         \label{degreeA}
     \end{subfigure}
     \hspace{-1mm}
     \begin{subfigure}[b]{.32\textwidth}
         \centering
         \includegraphics[width=0.75\textwidth]{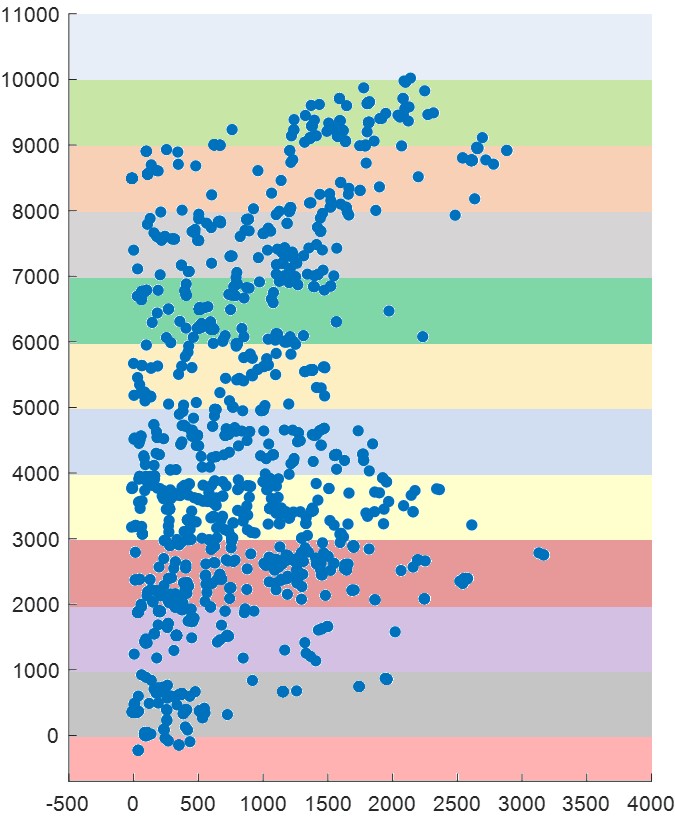}
         \caption{90$^{\circ}$}
         \label{degreeB}
     \end{subfigure}
     \hspace{-1mm}
     \begin{subfigure}[b]{.32\textwidth}
         \centering
         \includegraphics[width=0.75\textwidth]{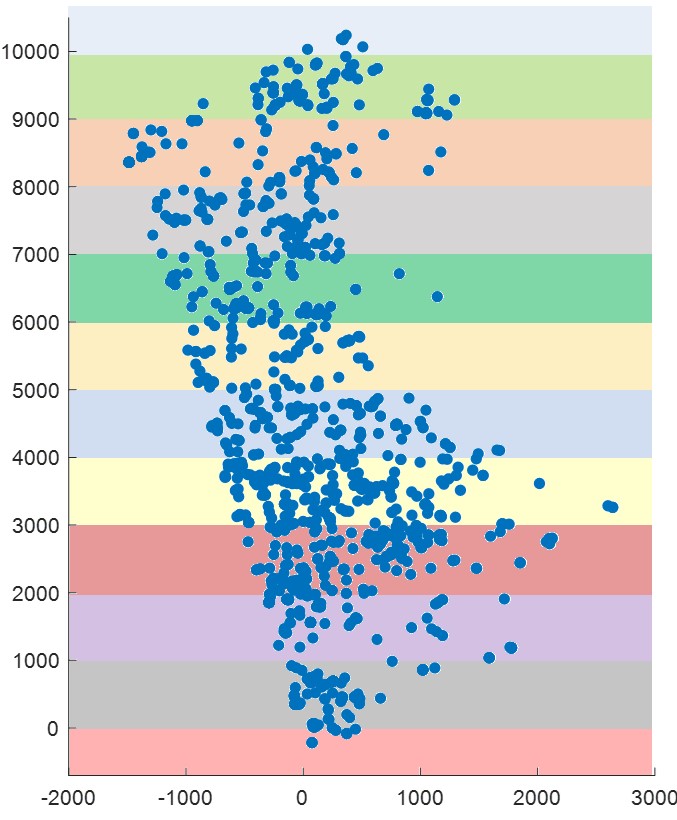}
         \caption{100$^{\circ}$}
         \label{degreeC}
     \end{subfigure}
     
        \caption{Node groups on sub axes according to the variation in the rotation angle of the main axis.}
        \label{degree}
\end{figure*}
%
\begin{figure*}[t!]
     \centering
     \begin{subfigure}[b]{.31\textwidth}
         \centering
         \includegraphics[width=0.95\textwidth]{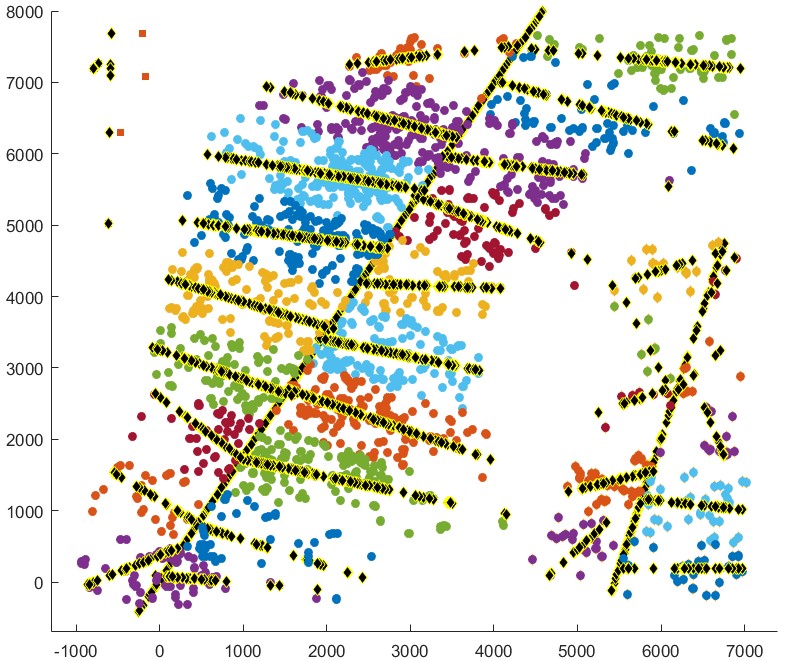}
         \caption{80$^{\circ}$}
         \label{axis2A}
     \end{subfigure}
     \hspace{-1mm}
     \begin{subfigure}[b]{.31\textwidth}
         \centering
         \includegraphics[width=0.95\textwidth]{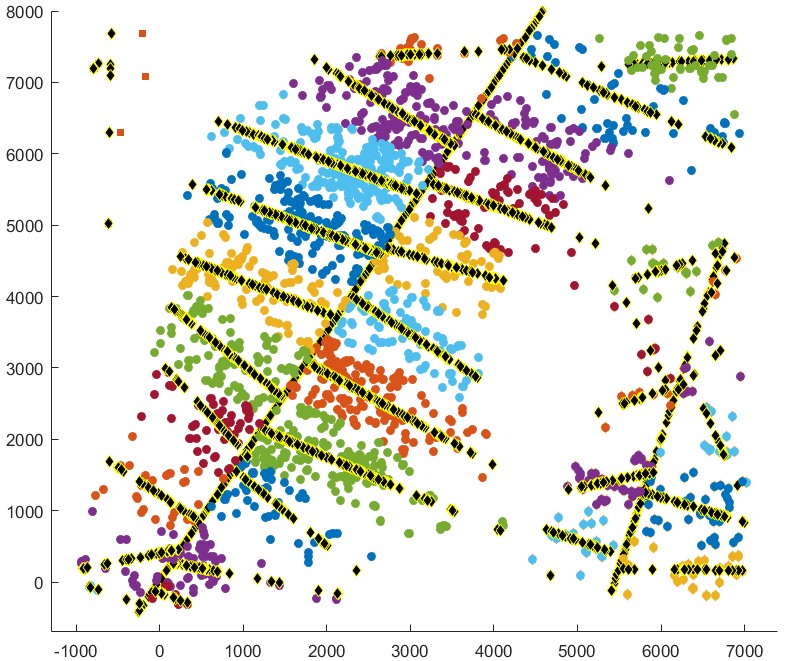}
         \caption{90$^{\circ}$}
         \label{axis2B}
     \end{subfigure}
     \hspace{-1mm}
     \begin{subfigure}[b]{.31\textwidth}
         \centering
         \includegraphics[width=0.95\textwidth]{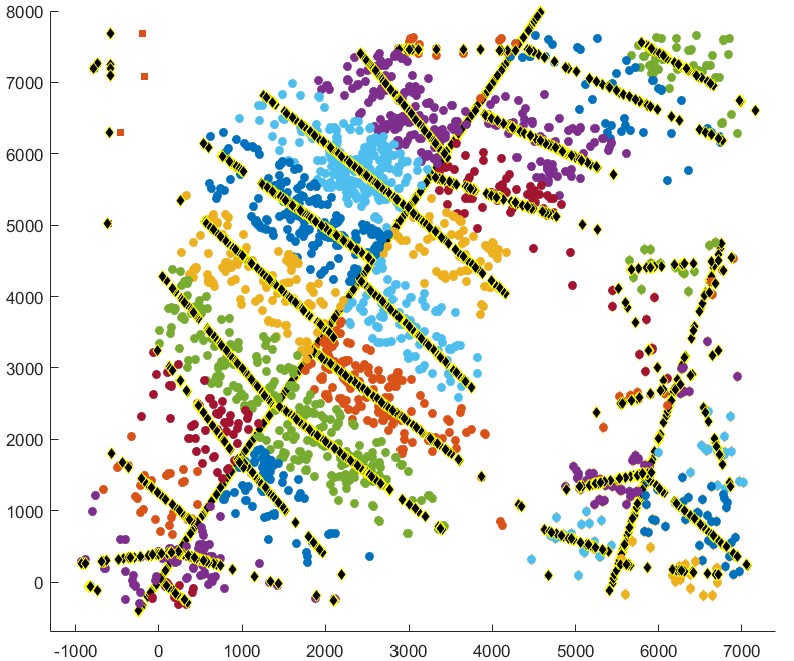}
         \caption{100$^{\circ}$}
         \label{axis2C}
     \end{subfigure}
 
        \caption{Creation of sub axes according to the variation in the rotation angle of the main axis.}
        \label{axis2}
\end{figure*}
%

\smallskip \noindent
\textbf{Relay device selection: }After the forwarding regions and sub axes are determined, we focus on selecting relay devices in order to forward messages in each region. 
\tcb{Along with Fig.~\ref{relaynode}, we elaborate on the relay-device selection step in each sub-axis region, where green devices serve as the selected relays and a red device is the destination.
As previously mentioned, Fig.~\ref{relaynode}(a) shows the set of partitioned forwarding regions according to the sub axes, each with twice the transmission range. Fig.~\ref{relaynode}(b) illustrates the set of selected relays for a certain partitioned forwarding region (i.e., a disjoint sub-area). In the sub-area, we select the first relay device such that the selected relay is not only farthest from the main axis but also nearest to the sub-axis among the relays within the transmission range, whose center is along the main axis, as depicted in Fig.~\ref{relaynode}(b). Then, we select the second relay device that is farthest from the main axis and nearest to the sub-axis out of all the relays within the transmission range, with a center at the first selected relay. This procedure is repeated until the destination is reached. This relay device selection enables us to significantly reduce the overlap of the transmission ranges of devices while enhancing the coverage probability. }

Motivated by our empirical finding that coverage probability depends heavily on the rotation angle of the main axis in each cluster (see Fig.~\ref{rotation_angle}), we introduce a method for searching for the rotation angle of the main axis, which enables us to create different sub axes.

Fig.~\ref{degree} shows the variation in the devices on the sub axes for the largest cluster in Fig.~\ref{clustering} when the rotation angles of the main axis are $80\,^{\circ}$, $90\,^{\circ}$, and $100\,^{\circ}$. \tcb{The different background colors represent the sub-axis groups according to the transmission range and rotation angle of the main axis.} Fig.~\ref{axis2} also shows how the sub axes are created according to the given rotation angles in Fig.~\ref{degree}. Now, we revisit Fig.~\ref{rotation_angle}, which shows the coverage probability versus rotation angle of the \tcb{ sub axes from the main axis}. The empirical findings reveal that the performance of the coverage probability can be significantly changed according to the values of the rotation angle. It is observed that the maximum coverage probability is obtained when the rotation angle is set to $83\,^{\circ}$, whereas a rotation angle of $107\,^{\circ}$ leads to the smallest coverage probability. These findings imply that the proper setting of the rotation angle of the main axis is crucial to guarantee satisfactory performance.
The overall procedure of the proposed \textsf{FiFo} method, which includes three phases, is summarized in Algorithm 1.

\begin{figure}[!t]
\centering
\includegraphics[width=0.99\columnwidth]{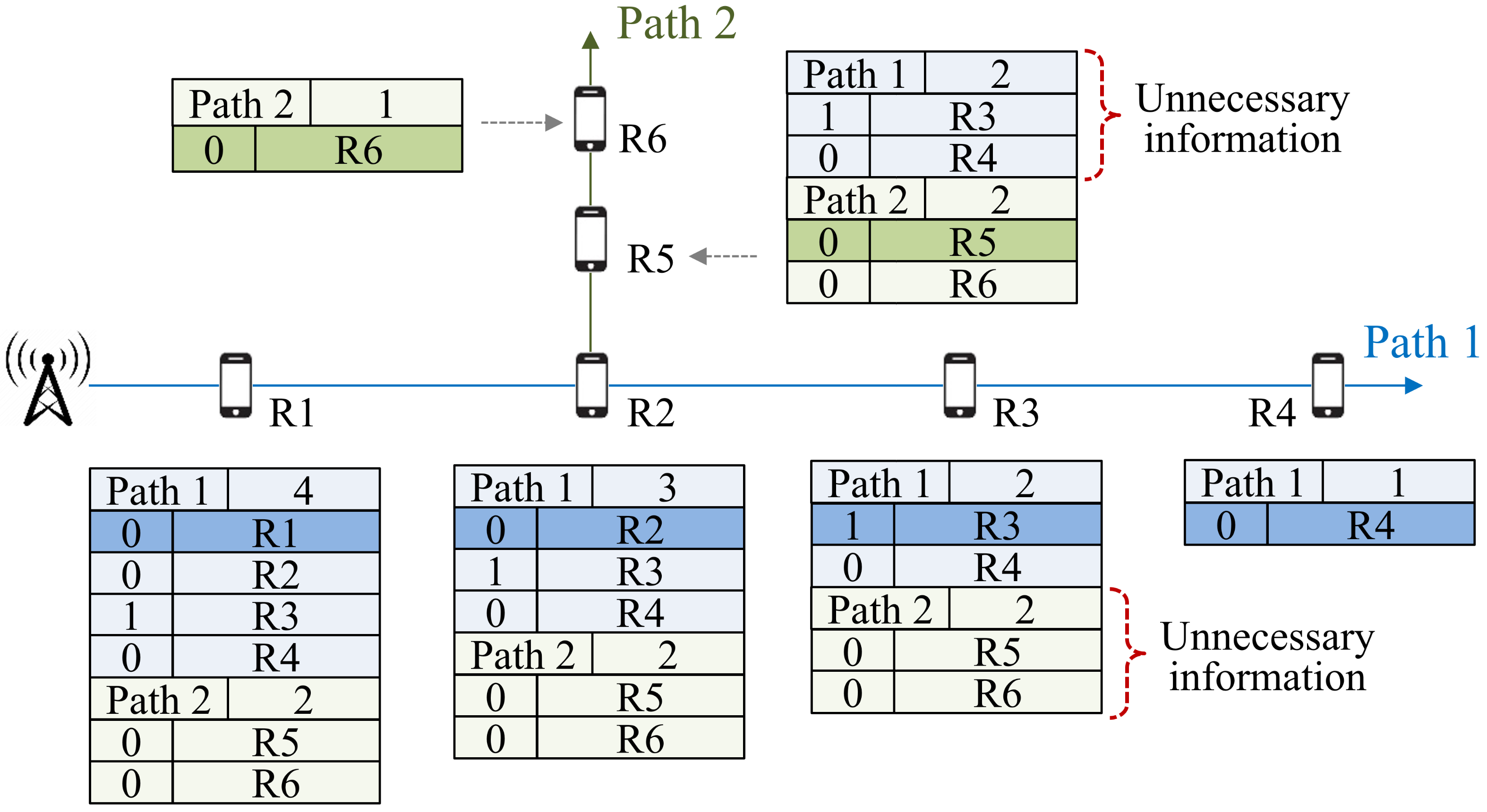}
\caption{\jskim{An example of data dissemination based on the \textsf{FiFo}-enabled message format.}}
\label{fig:format_example}
\end{figure}

\jskim{
\subsection{Fishbone-Shaped Data Dissemination}
    To disseminate a message along the fishbone-shaped forwarding paths, we explain how relay selection information is forwarded while each device determines whether or not to relay.
    %
    We first describe a new message format for \textsf{FiFo} and then explain the detailed procedure by which the message is propagated.
    As shown in Fig.~\ref{fig:message_format}, the \textsf{FiFo}-enabled message contains the following fields.
\begin{itemize}
        \item Source Device ID: Identification of the sender (i.e. the previous relay device);
        \item Message Sequence Number: A unique number to identify the message;
        \item Number of Paths: The total number of the main and sub axes;
        \item Path Information: The field containing information about relay devices in each forwarding path.
\end{itemize}
        The number of path information fields is determined by the number of paths and 
    each piece of path information includes the following specific fields:
    \begin{itemize}
        \item Path ID: Identification of the path;
        \item Number of Relays: The number of relay devices in the corresponding path;
        \item Relay Flag: A flag indicating whether it is a relay device that has passed a path junction;
        \item Relay Device ID: Identification of the relay device in the corresponding path.
\end{itemize}
\begin{figure}[!t]
\centering
\includegraphics[width=0.99\columnwidth]{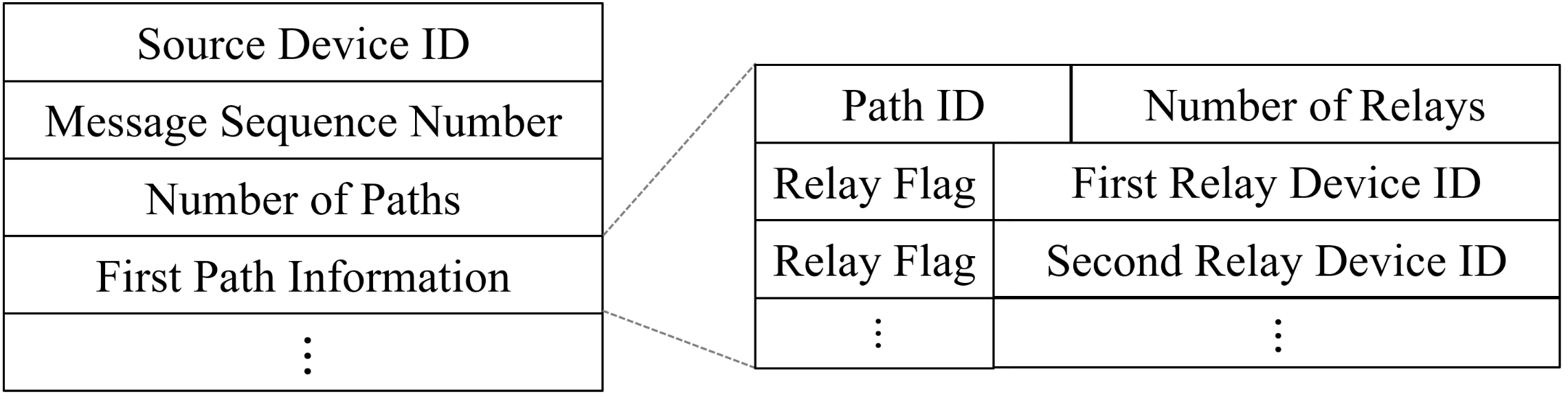}
\caption{\jskim{\textsf{FiFo}-enabled message format.}}
\label{fig:message_format}
\end{figure}
     The source device ID and message sequence number are used to check for message duplication. When receiving a message, each device verifies its first relay device ID in the first-path information field. If the relay device ID is matched, then the device sends a message to its neighbor as a relay. 
    In contrast, if the relay flag is marked, then the device checks for another relay device ID, which is the first relay node ID of the following path information field.
    If the relay device ID is matched, then the device forwards the message as the first device that starts relaying along another path. When the relay flag is not marked, after excluding its own ID from the path information and updating the field value (the number of relays), the device selected as a relay forwards the message. In contrast, if the relay flag is marked, then the forwarding path information to be delivered varies significantly depending on the path to which the relay device belongs. An example of this procedure is illustrated in Fig.~\ref{fig:format_example}, where there are two paths including one main axis and one sub axis.
}
\section{Dataset Description} 
\label{sec_dataset}

In this section, we elaborate on our real-world dataset, which includes the network topology. The distribution of IoT devices (or users equipped with IoT devices) in two-dimensional wireless networks is typically modeled by a homogeneous Poisson point process (PPP) owing to its analytical tractability~\cite{Weberetal}. However, user distribution is not completely spatially random, that is, users are either clustered or sometimes more regularly distributed. Clustered users may occur owing to geographical factors, especially in large-scale (e.g., city-scale or borough-scale) networks; for example, in a social gathering, people tend to cluster into small groups, while in a downtown area, users are clustered indoors such as in buildings, and in a cognitive network, active cognitive users tend to be clustered. Thus, the location of devices can also be modeled as a stationary and isotropic Poisson cluster process~\cite{Stoyanetal}. Because the performance of our data dissemination depends heavily on user distribution, it is imperative to use real geolocation information collected from mobile phones and online social network data.

In our study, to evaluate the performance of our data dissemination method (i.e., \textsf{FiFo}), we use a Twitter dataset, previously used in~\cite{Shinetal} and collected globally via the Twitter Streaming Application Programming Interface (API) from September 22, 2014 to October 23, 2014 (about one month). This is because Twitter~\cite{Kwaketal} is currently one of the most popular microblogs (or social media), and at the start of 2015, it played a vital role in facilitating social contacts, boasting 284 million active users per month, publishing 500 million tweets daily from their web browsers and smartphones. It was found that the streaming API returns an almost complete set of {\em geo-tagged} tweets matching a query provided by the streaming API user despite sampling. 

\begin{figure*}[!t]
     \centering
     \begin{subfigure}[t]{.305\textwidth}
         \centering
         \includegraphics[width=\textwidth]{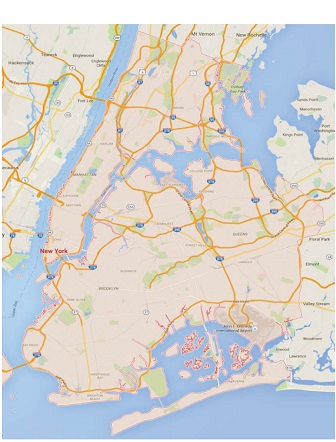} 
         \caption{New York City in the US}
         \label{addingARNA}
     \end{subfigure}
     \hspace{-1mm}
     \begin{subfigure}[t]{.32\textwidth}
         \centering
         \includegraphics[width=\textwidth]{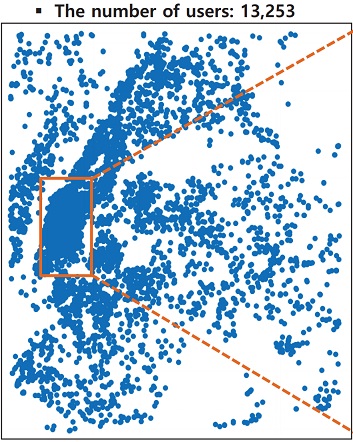} 
         \caption{New York City (as whole city)}
         \label{addingARNB}
     \end{subfigure}
     \hspace{-1mm}
     \begin{subfigure}[t]{.32\textwidth}
         \centering
         \includegraphics[width=\textwidth]{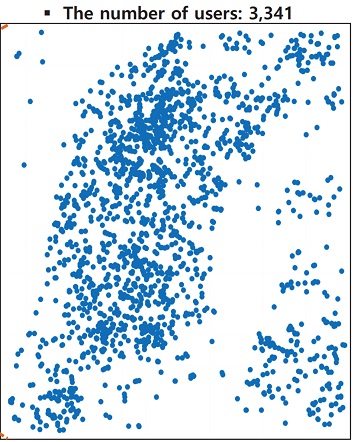}
         \caption{Down-town region in New York City ($8km^{2}$)}
         \label{addingARNC}
     \end{subfigure}
         \caption{\textcolor{blue}{IoT-}device distribution over two query regions in New York City.}
         \hrulefill
        \label{Dataset}
\end{figure*}

From the entire dataset in~\cite{Shinetal}, through query processing with geographic bounding boxes (expressed as both latitude and longitude), we then obtained the filtered geo-tagged tweets that were posted only in New York City (the most populous city in the US) and a certain downtown region in the city with an area of 8km$^2$, as illustrated in Fig.~\ref{Dataset}. These two regions were selected because they provide a high population density and high Twitter popularity, which fit into data dissemination environments. Thereafter, from the above-filtered tweets, we focused only on four-hour records that took place between 0 am and 4 am UTC/GMT. These short-term (four hours) records are sufficient to examine how data are disseminated via short-range D2D communications.

The statistics of the used Twitter data are summarized in Table~\ref{stats}. In this dataset, each tweet record has a geo-tag and a timestamp indicating where, when, and by whom the tweet was sent. The location information provided by the geo-tag is denoted by latitude and longitude, which are measured in degrees, minutes, and seconds.

\begin{table}[!t]
\centering \caption{Statistics of the dataset: the number of Twitter users} \label{stats} \hspace{1cm}
\begin{tabular}{lllll}
\cline{1-2} \multicolumn{1}{|l|}{\bf{Region}}           & \multicolumn{1}{c|}{\bf{Number of users}}   & \\
\cline{1-2} \multicolumn{1}{|l|}{New York City (as whole city)}         & \multicolumn{1}{c|}{{\begin{tabular}[c]{@{}l@{}}{13,253}\end{tabular}}} & \\
\cline{1-2} \multicolumn{1}{|l|}{Down-town region in New York City (8km$^2$)} & \multicolumn{1}{c|}{\begin{tabular}[c]{@{}l@{}}{3,341}\end{tabular}} & \\
\cline{1-2} & & & &
\end{tabular}
\end{table}

We examine data from all possible devices (sources) that indicate the user's location at the time that they access Twitter. The statistics based on our dataset demonstrate that a large majority of Twitter users in our sample posted geo-tagged tweets through smartphones rather than web browsers on a desktop or laptop computer.
This reveals that our dataset is much more inclined towards geo-tagged tweets transmitted through GPS interface-enabled IoT devices.

Each tweet contains several entities that are distinguished by their respective fields. For our data dissemination analysis, we adopt the following four essential fields from the data of the tweets:
\begin{itemize}
\item {\em user\_id\_str}: string representation of the sender ID
\item {\em lat}: latitude of the sender
\item {\em lon}: longitude of the sender
\item {\em created\_at}: UTC/GMT time when the mention is delivered, i.e., the timestamp
\end{itemize}
Note that the two location fields, {\em lat} and {\em lon}, correspond to spatial (geo-tagged) information, whereas the last field, {\em created\_at}, represents temporal (time-stamped) information.

\section{Numerical Evaluation} 
\label{sec_results}

In this section, we first present four benchmark data dissemination methods for comparative studies. After describing our performance metrics and experimental settings, we comprehensively evaluate the performance of our \textsf{FiFo} method and the four benchmark methods. In addition, we evaluate the runtime complexity of all the methods.

\subsection{Benchmark Methods} 
\label{SEC:Benchmark}

\tcb{Detailed descriptions of the four benchmark data dissemination methods in massive IoT networks are as follows.}


\begin{itemize}
    \item {\bf Epidemic routing~\cite{epidemic}:} This benchmark method performs omni-directional contact-based diffusion to improve network connectivity. The epidemic routing rapidly distributes messages over the underlying wireless network at the cost of excessive overlaps of transmission ranges of devices.
    
    \item {\bf Modified broadcast incremental power (BIP)~\cite{BIP}:} Another benchmark method for alleviating the excessive, unnecessary message transmissions/receptions is BIP, which is built on a spanning tree rooted at a given device. Its computational overhead is relatively heavy owing to the broadcast tree construction process. Note that, unlike the original BIP algorithm in~\cite{BIP}, the modified BIP algorithm used in this study does not perform power control. 
    
    \item {\bf Probabilistic flooding (PF)~\cite{BroadcastStorm,BulkDataDissemination}:} In probabilistic flooding, upon receiving a new message for the first time, each device (receiver) rebroadcasts the message once with a certain probability $p_f$ (which will be specified later). This results in a significant reduction in the number of receptions compared to the epidemic forwarding. In our experiments, as each device determines whether to forward the received message randomly, we run this algorithm $5\times 10^4$ times for a given setting to calculate the average performance.

    \item {\tcb{\textbf{Neighbor-based probabilistic broadcast (NPB)~\cite{NPB}: } This is a probabilistic broadcast algorithm based on the knowledge of neighboring devices that determines a rebroadcast probability. To calculate the rebroadcast probability, NPB uses the number of uncovered neighbor devices along with an additional coverage ratio and an adaptive connectivity factor. As a result, according to the device density and distributions, NPB adaptively obtains the rebroadcast probability of each device.
    %
    }}
    

\end{itemize}




\begin{figure*}[!t]
     \centering
     \begin{subfigure}[t]{.32\textwidth}
         \centering
         \includegraphics[width=\textwidth]{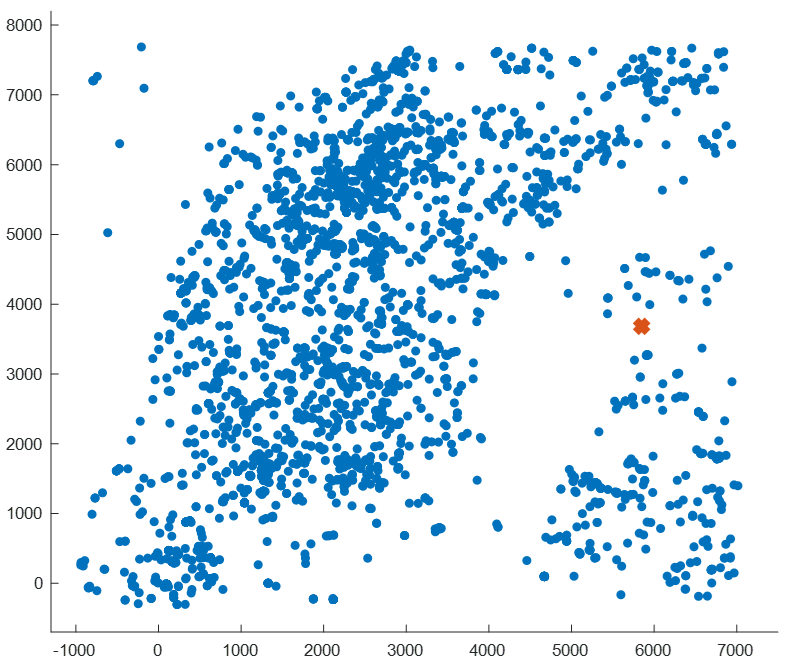}
         \caption{Initial network deployment of an additional relay device (marked with a red cross (`\textcolor{red}{$\times$}'))}
     \end{subfigure}
     \hspace{-1mm}
     \begin{subfigure}[t]{.32\textwidth}
         \centering
         \includegraphics[width=\textwidth]{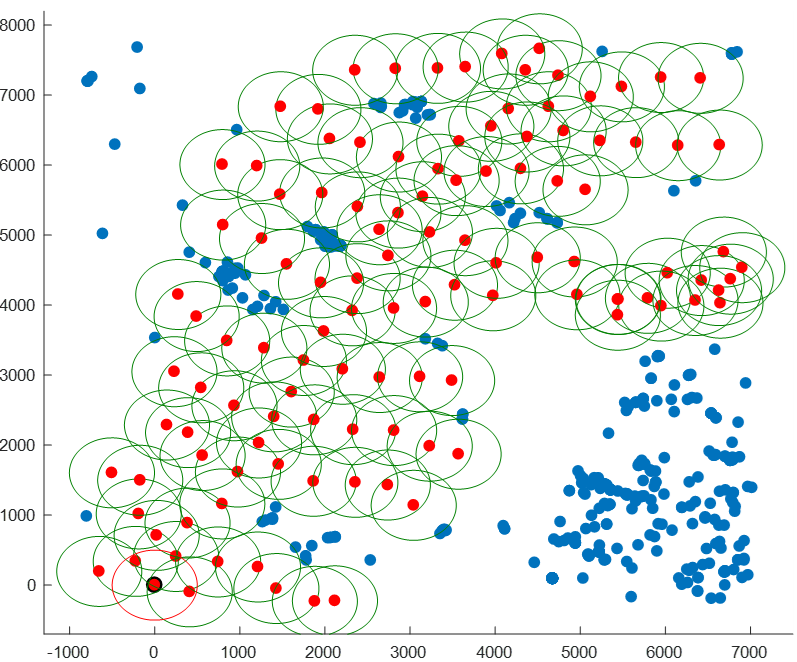}
         \caption{Data dissemination without the relay deployment, where devices that have received no message are marked with a blue circle (`\textcolor{blue}{$\bullet$}')}
     \end{subfigure}
     \hspace{-1mm}
     \begin{subfigure}[t]{.32\textwidth}
         \centering
         \includegraphics[width=\textwidth]{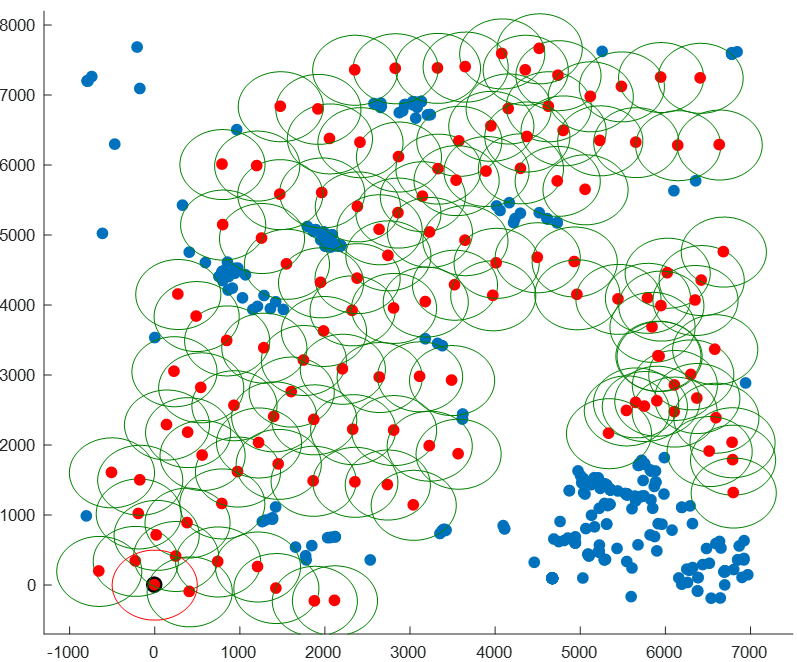}
         \caption{Data dissemination with the relay deployment, where devices that have received no message are marked with a blue circle (`\textcolor{blue}{$\bullet$}')}
     \end{subfigure}
 
        \caption{Data dissemination map before and after performing our \textsf{FiFo} method by deploying an additional relay device at $(5848.28, 3682.84)$ over the query region.}
        \label{addingARN}
\end{figure*}

\subsection{Key Performance Metric}
To empirically validate the performance of the proposed \textsf{FiFo} method over the above four benchmark methods, we access the performance via a new metric called {\em forwarding efficiency}, which is appropriate for the performance evaluation of data dissemination. 

The forwarding efficiency $\eta$ is defined as the ratio of the coverage probability to the average number of transmissions per device, and is expressed as
\begin{equation}\label{objective1}
\eta = \frac{\sum_{i \in \mathcal{N}}{\mathcal{I}(i)}}{\sum_{i \in
\mathcal{N}}{\mathcal{F}(i)}},
\end{equation}
where $\mathcal{N}$ is the set of IoT devices in the underlying network, $\mathcal{F}(i)$ is the number of transmissions at device $i$, and $\mathcal{I}(i)$ is an indicator function. If device $i$ receives a message successfully, $\mathcal{I}(i) = 1$; otherwise, $\mathcal{I}(i) = 0$. The coverage probability is calculated as $\frac{\sum_{i \in \mathcal{N}}{\mathcal{I}(i)}}{|\mathcal{N}|}$, where $|\mathcal{N}|$ is the cardinality of $\mathcal{N}$.

\subsection{Experimental Settings}
In this subsection, the experimental setup is described. As specified in Section~\ref{sec_dataset}, we use the Twitter dataset collected from a certain downtown region in New York City to model a network environment in which devices (users) are clustered (see Fig.~\ref{Dataset} for device distribution). Because allowing a sufficiently large number of hops through relaying devices is not feasible, especially for delay-sensitive applications, such as emergency and disaster alarms, we limit the number of maximum allowable hops for each message to a certain value. In other words, if the running algorithm exceeds the maximum number of hops, it is terminated. In our experiments, the maximum number of hops is set to 25; however, it can also be set to a different value, depending on the circumstances.

\subsection{Experimental Results}
Our empirical study is designed to answer the following \tcb {four} research questions (RQs).

\begin{itemize}
\item {\em RQ1.} How does the location of a source device affect the performance?
\item {\em RQ2.} How does adding an additional relay device influence the performance?
\item \tcb{{\em RQ3.} How much does \textsf{FiFo} improve the energy efficiency compared to existing studies?
}
\item \tcb{{\em RQ4.} How does the combination of the main axis of \textsf{FiFo} and existing studies influence performance?
}
\end{itemize}

To answer these research questions, we comprehensively perform the following experiments.

\subsubsection{\textbf{Location of a Source Device (RQ1)}}
In our experiments, we consider two scenarios: the fixed and random locations of a source device. For the fixed case, as shown in Fig.~\ref{addingARN}, a device located near the main axis is selected as a source such that our \textsf{FiFo} method disseminates messages quite effectively along the created main axis and sub-axes. \tcb{In Fig.~\ref{addingARN}(a), a device with a red cross (`\textcolor{red}{$\times$}') represents an additional relay device. In Figs. \ref{addingARN}(b) and \ref{addingARN}(c), devices with a red circle (`\textcolor{red}{$\bullet$}') and devices with a blue circle (`\textcolor{blue}{$\bullet$}') represent devices that have relayed a message and no message, respectively.} The (relative) $x$- and $y$-coordinates of the source device is set to $(0, 0)$ (see Fig.~\ref{addingARN}(a)). However, for the random case, a source device is randomly selected for each experiment. We run each algorithm $5\times10^4$ times while randomly changing the location of the source device.

A performance comparison between our \textsf{FiFo} method and four benchmark message forwarding methods in Section~\ref{SEC:Benchmark}, including epidemic routing, modified BIP, and probabilistic flooding, is presented in Table~\ref{Table:results1} with respect to the forwarding efficiency, where the results for two source device deployment scenarios are shown. From Table~\ref{Table:results1}, our findings are as follows.

\begin{itemize}
    \item The \textsf{FiFo} method significantly and consistently outperforms all benchmark methods in terms of the forwarding efficiency. The maximum improvement rate of our \textsf{FiFo} method ($X$) over the second-best performer ($Y$) (i.e., modified BIP) is $79.08\%$, where the improvement rate (\%) is given by $\frac{X-Y}{Y}\times100$.
    \item For all methods, only a marginal gain in performance is achieved when a fixed source location is used over its counterpart (i.e., a random source location). This, however, does not apply to the probabilistic flooding results ($p_f=0.1$). This implies that the location of a source device does not play a crucial role in determining the performance of the message forwarding method.
    \item Epidemic routing exhibits the worst performance among the four methods. This is because epidemic routing is inherently designed to enhance the network connectivity at the cost of high overlaps of transmission ranges of devices, thus resulting in a large number of transmissions per device.
    \item Likewise, as the forwarding probability $p_f$ increases, the performance of probabilistic flooding gets degraded due to the increment in the number of message transmissions/receptions.
    \item \tcb{Although NPB is a topology-aware data dissemination method, it still has practical challenges in discovering both central data dissemination spine and relay devices. Thus, the improvement of the forwarding efficiency $\eta$ is not significantly large compared to PF.}
\end{itemize}

\begin{table}[!t]
\centering 
\caption{Performance comparison of \textsf{FiFo} and benchmark algorithms in terms of $\eta$ under the fixed and random locations of a source device.
} 
\begin{tabular}{|c|c|c|}
\hline 
Method & Fixed location &  Random location \\
\hhline{|t=:t=:t=|} 
Epidemic routing~\cite{epidemic} & 0.0124  & 0.0125\\
\hline
PF ($p_f=0.1$)~\cite{BulkDataDissemination} & 0.0591 & 0.0756 \\
\hline
PF ($p_f=0.5$)~\cite{BulkDataDissemination} & 0.0208 & 0.0224 \\
\hline
\tcb{NPB}~\cite{NPB} & \tcb{0.1213} & \tcb{0.1208} \\
\hline
Modified BIP~\cite{BIP} & 0.1984 & 0.1723 \\
\hline
\textsf{FiFo} (Proposed) & 0.3553 & 0.3089\\
\hline
\end{tabular} \label{Table:results1}
\end{table}

\subsubsection{\textbf{Deployment of an Additional Relay (RQ2)}}

Although our \textsf{FiFo} method significantly outperforms benchmark methods in terms of forwarding efficiency, it is likely that the transmission range covered by IoT devices may be limited because of potential coverage holes, as illustrated in Fig.~\ref{addingARN}(b). Thus, a non-negligible number of devices may not be able to receive the disseminated messages. To overcome this problem, we consider an additional relay, such as a movable relay device or an unmanned aerial vehicle (UAV), as a replenishment step built upon the \textsf{FiFo} method, which enables us to forward the messages to devices within each coverage hole. In our experiments, we deployed this additional relay at $(5848.28, 3682.84)$, corresponding to the $x$- and $y$-coordinates, to effectively serve the devices that are within the coverage hole existing in the lower right area (see Fig.~\ref{addingARN}(b)), where the source device is assumed to be fixed at $(0, 0)$ (i.e., a fixed source location is assumed). Figs.~~\ref{addingARN}(b) and \ref{addingARN}(c) illustrate the data dissemination map before and after performing the \textsf{FiFo} method with additional relay deployment, respectively.

\begin{table}[!t]
\centering 
\caption{Comparison of the performances of \textsf{FiFo} without deploying an additional relay device and benchmark algorithms in terms of $\frac{\sum_{i \in \mathcal{N}}{\mathcal{I}(i)}}{|\mathcal{N}|}$ and $\eta$ when the location of the source device is fixed.
} 
\begin{tabular}{|c|c|c|}
\hline 
Method & Coverage prob. & Forwarding eff. \\
\hhline{|t=:t=:t=|} 
Epidemic routing~\cite{epidemic} & 0.8982 & 0.0117\\
\hline
PF ($p_f\!=\!0.1$)~\cite{BulkDataDissemination} &0.6677 & 0.0624 \\
\hline
PF ($p_f\!=\!0.5$)~\cite{BulkDataDissemination} & 0.8751 & 0.0208 \\
\hline
\tcb{NPB}~\cite{NPB} & \tcb{0.8375} & \tcb{0.1215} \\
\hline
Modified BIP~\cite{BIP} & 0.7971 & 0.1960 \\
\hline
\textsf{FiFo} (Proposed) & 0.8521 & 0.3421 \\
\hline
\end{tabular} \label{Table:results2}
\end{table}

A performance comparison between our \textsf{FiFo} method and four benchmark message forwarding methods is presented in Table~\ref{Table:results2} with respect to the coverage probability and forwarding efficiency, where the results with and without the additional relay are shown. From Table~\ref{Table:results2}, the following observations can be made.

\begin{itemize}
    \item For the case in which an additional relay is deployed, the \textsf{FiFo} method also significantly outperforms all benchmark methods in terms of the forwarding efficiency. In this case, the improvement rate of \textsf{FiFo} over the second best performer (i.e., modified BIP) is $74.54\%$ in terms of the forwarding efficiency.
    \item When our \textsf{FiFo} method is employed, the forwarding efficiency gap between the cases with and without the relay deployment is very marginal, while the enhancement of the coverage probability is quite remarkable by adding only one additional relay.
    \item Epidemic routing exhibits the highest coverage probability; however, the gain over \textsf{FiFo} is quite negligible. 
    \item In probabilistic flooding, as the forwarding probability $p_f$ increases, the coverage probability is largely improved, which is rather obvious.
\end{itemize}


\subsubsection{\textbf{\tcb{Energy Efficiency (RQ3)}}}
\tcb{
To show the energy consumption characteristics of each algorithm, we present energy efficiency $\alpha$ as another performance metric,
%
which is defined as the ratio of the coverage probability to the total amount of power consumed by all devices and is expressed as
\begin{equation}\label{eq_ee}
\alpha = \frac{\sum_{i \in
\mathcal{N}}{\mathcal{I}(i)}}{\frac{P_t}{\beta}\sum_{i \in
\mathcal{N}}{\mathcal{F}(i)}+P_r\sum_{i \in
\mathcal{N}}{\mathcal{R}(i)}+P_s\sum_{i \in
\mathcal{N}}{\mathcal{U}(i)}},
\end{equation}
where $\beta$ is the power amplifier efficiency; $P_t$, $P_r$, and $P_s$ are the transmit power, receive power, and standby power, respectively, per device; $\mathcal{F}(i)$ is the number of transmissions at device $i$; $\mathcal{R}(i)$ is the number of receptions at device $i$; and $\mathcal{U}(i)$ is the number of devices that receive no message sent by device $i$. Similar to ~\cite{energyefficiency,3gpp}, we set $P_t$, $P_r$, and $P_s$ to 480mW, 75mW, and 0.015mW, respectively.
}
\begin{figure}[!t]
\centering
\includegraphics[width=7cm]{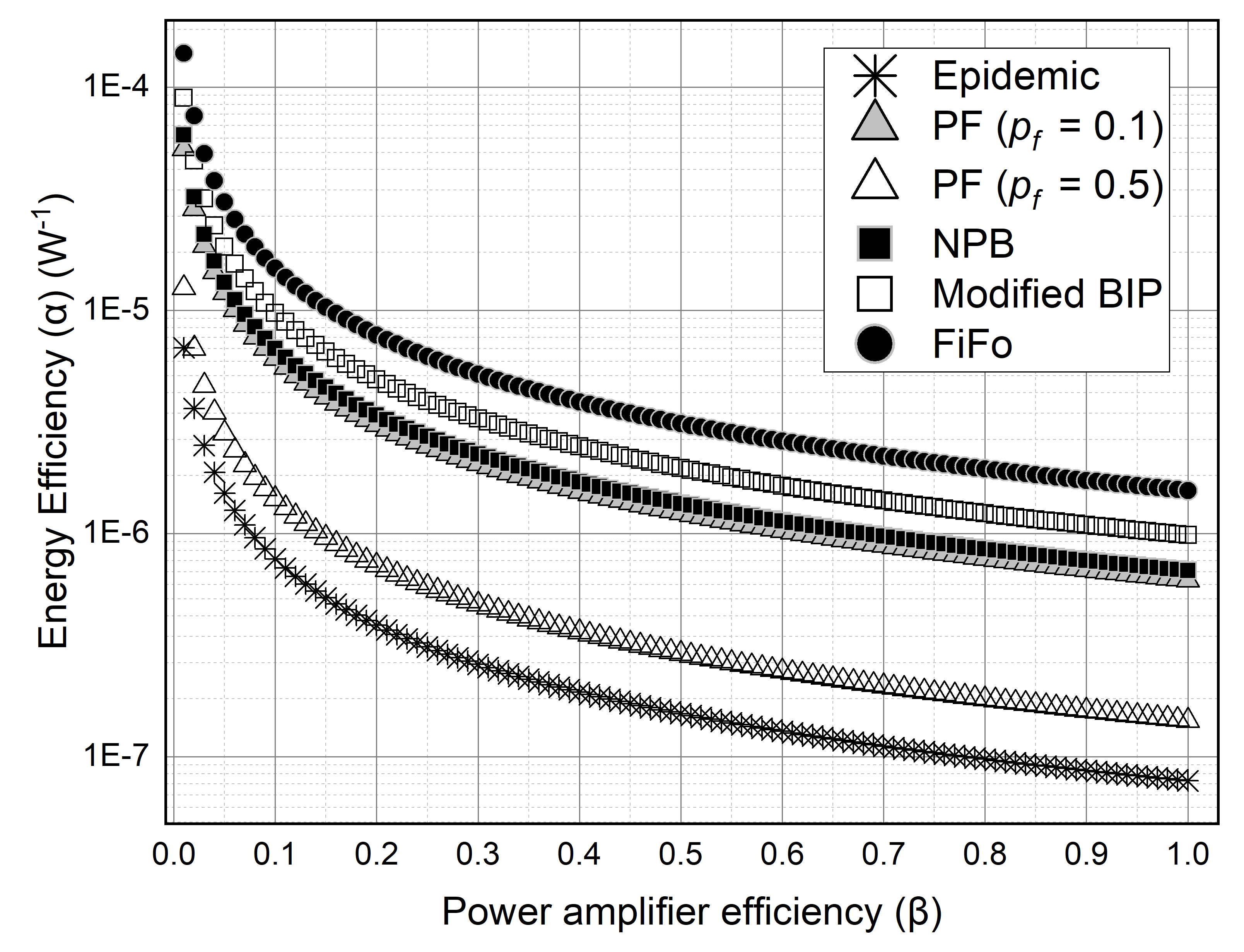}
\caption{\tcb{Energy efficiency $\alpha$ of \textsf{FiFo} and four benchmark methods according to different values of the power amplifier efficiency $\beta$, where the fixed location of a source device and no deployment of an additional relay device are assumed.}}
\label{ee}
\end{figure}

\tcb{Fig.~\ref{ee} shows the energy efficiency $\alpha$ ($W^{-1}$) of \textsf{FiFo} and the benchmark message forwarding methods according to different values of $\beta\in[0,1]$ when the fixed location of a source device and no deployment of an additional relay device are assumed. Our findings are as follows.
\begin{itemize}
    \item The proposed \textsf{FiFo} consistently outperforms all benchmark methods based on the overall values of $\beta$.
    \item Unlike the case of the forwarding efficiency, the performance of PF ($p_f =0.1$) is superior to that of PF ($p_f =0.5$). This is because PF ($p_f =0.1$) utilizes a less number of relay devices, thus resulting in a lower number of receptions.
\end{itemize}
}  


\subsubsection{\textbf{\tcb{\textsf{FiFo} Variants (RQ4)}}}
\tcb{To compare the performance of \textsf{FiFo} with various topology-aware broadcasting strategies exploiting the status of neighboring devices, we employ several \textsf{FiFo} variants. In Tables~\ref{Table:fifo_ma1} and \ref{Table:fifo_ma2}, \textsf{FiFo-MA} represents the main axis creation component of \textsf{FiFo}. Built upon \textsf{FiFo-MA}, we make two variants in combination with PF and NPB. Although these variants exploit the main axis creation part of \textsf{FiFo}, we can see that many limitations still exist in choosing relay devices that disseminate messages. Our empirical evaluation clearly demonstrates that the performance of the proposed \textsf{FiFo} is much superior to that of its variants.
}

\begin{table}[!t]
\centering 
\caption{\tcb{
Performance comparison of \textsf{FiFo} and its variants in terms of $\eta$ for fixed and random locations of the source device.
}}
\label{Table.withFiFo} 
\begin{tabular}{|c|c|c|}
\hline 
Method & Fixed location &  Random location \\
\hhline{|t=:t=:t=|} 
\tcb{\textsf{FiFo-MA} with PF ($p_f\!=\!0.1$)} & \tcb{0.0759} & \tcb{0.0758}\\
\hline
\tcb{\textsf{FiFo-MA} with PF ($p_f\!=\!0.5$)} &  \tcb{0.0209}  & \tcb{0.0209} \\
\hline
\tcb{\textsf{FiFo-MA} with NPB} & \tcb{0.1218} & \tcb{0.1198} \\
\hline
\textsf{FiFo} (Proposed) & 0.3553 & 0.3089\\
\hline
\end{tabular} \label{Table:fifo_ma1}
\end{table}

\begin{table}[!t]
\centering 
\caption{\tcb{Performance comparison of \textsf{FiFo} and its variants in terms of $\frac{\sum_{i \in \mathcal{N}}{\mathcal{I}(i)}}{|\mathcal{N}|}$ and $\eta$ for the case where the location of the source device is fixed.
} }
\begin{tabular}{|c|c|c|}
\hline 
Method & $\frac{\sum_{i \in \mathcal{N}}{\mathcal{I}(i)}}{|\mathcal{N}|}$ &  $\eta$ \\
\hhline{|t=:t=:t=|} 
\tcb{\textsf{FiFo-MA} with PF ($p_f\!=\!0.1$)} & \tcb{0.7984} & \tcb{0.0759} \\
\hline
\tcb{\textsf{FiFo-MA} with PF ($p_f\!=\!0.5$)} & \tcb{0.8793} & \tcb{0.0208} \\
\hline
\tcb{\textsf{FiFo-MA} with NPB} & \tcb{0.8629} & \tcb{0.1213} \\
\hline
\textsf{FiFo} (Proposed) & 0.8521 & 0.3421 \\
\hline
\end{tabular} \label{Table:fifo_ma2}
\end{table}

\subsection{Empirical Evaluation of Complexity}
\begin{figure}[!t]
\centering
\includegraphics[width=9cm]{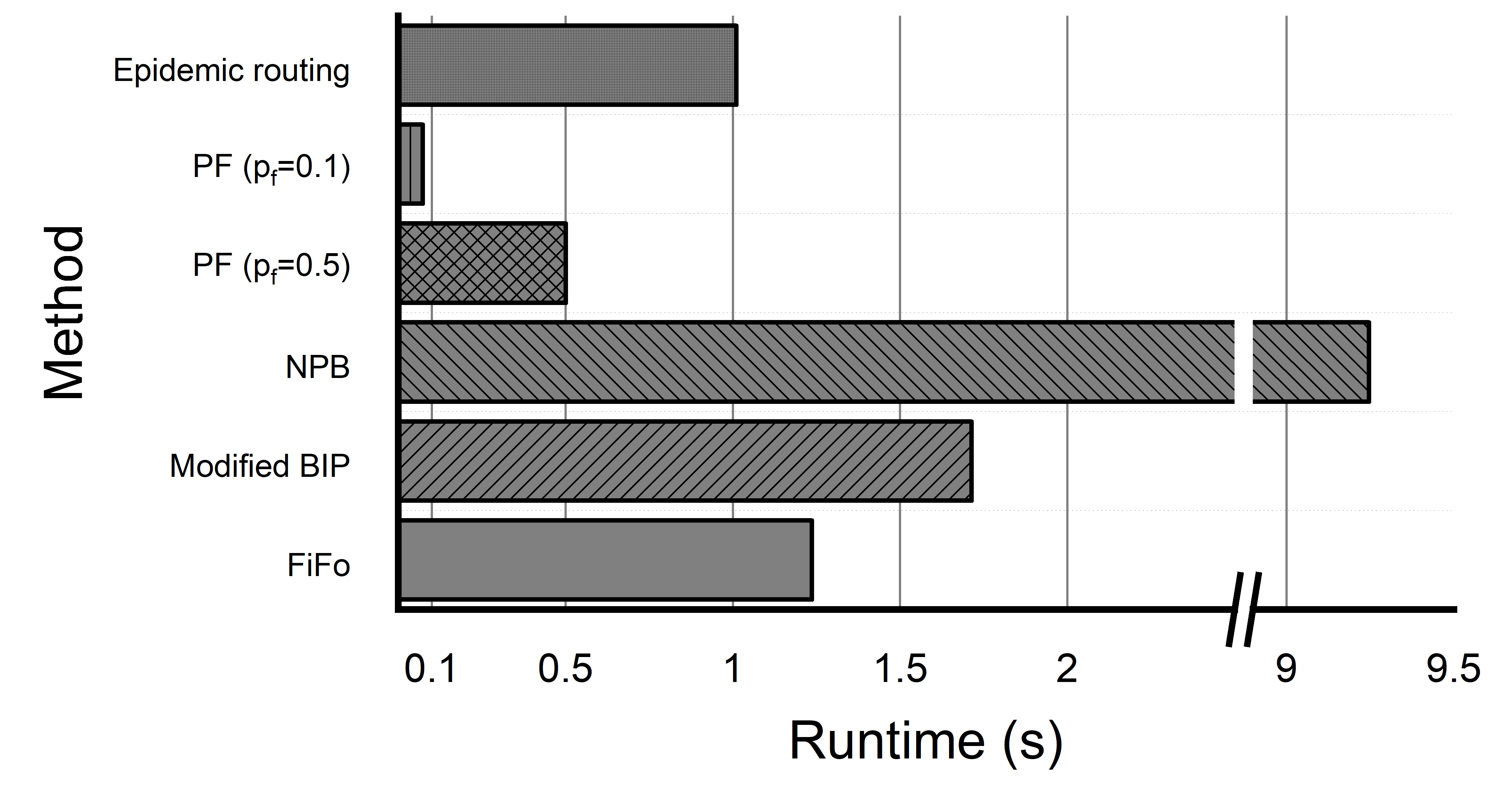}
\caption{\tcb{Runtime complexity (in seconds) of the \textsf{FiFo} and four benchmark methods, where a fixed location of the source device and no deployment of a relay device are assumed.}}
\label{cputime}
\end{figure}

In this subsection, we empirically demonstrate runtime complexity via experiments. A fixed location of the source device and no deployment of an additional relay are assumed for the complexity evaluation. In Fig.~\ref{cputime}, we show the runtime complexity in seconds when the \textsf{FiFo} and \tcb{four} benchmark methods, including epidemic routing, probabilistic flooding (with $p_f=0.1, 0.5$), \tcb{NPB}, and modified BIP, are executed. For probabilistic flooding, because we run the algorithm $5\times 10^4$ times for a given experimental setting, we measure the average runtime. From Fig.~\ref{cputime}, it is observed that the \textsf{FiFo} method requires only $66.64\%$ of the runtime required by the modified BIP (the second-best performer) while exhibiting a much higher forwarding efficiency than the modified BIP. In addition, the runtime complexity of epidemic routing and \textsf{FiFo} is comparable; nevertheless, the gain of \textsf{FiFo} over epidemic routing in terms of forwarding efficiency is significant (refer to Tables~\ref{Table:results1} and~\ref{Table:results2}).

\section{Related Work}
\label{sec_model}

The method proposed in this study is related to data-dissemination approaches in wireless ad hoc networks.


{\bf Data dissemination in static networks.} The study of data dissemination, also known as message forwarding or flooding, in wireless ad hoc networks where {\em static} devices are deployed, has been steadily conducted in the literature. A well-known algorithm called epidemic routing~\cite{epidemic} was presented to improve network connectivity, particularly in desert, sea, and disaster environments. In~\cite{epidemic}, through multiple handshaking steps for message exchanges among nodes, messages were rapidly distributed over the underlying network at the cost of excessive overlaps of transmission ranges of nodes owing to omnidirectional contact-based diffusion. As an alternative, BIP~\cite{BIP}, which alleviates excessive and unnecessary message transmissions/receptions, was proposed for message forwarding. Because the BIP algorithm is built upon a spanning tree rooted at a given node, it incurs heavy computational overhead when constructing the broadcast tree. In addition, in delegation forwarding~\cite{DelegationForwarding}, a node performs message forwarding only to a neighboring node with the best quality metric, which ensures satisfactory performance with a low transmission cost. However, there is still a practical challenge in developing the delegation forwarding algorithm because the distribution of node qualities is usually unknown and its message delivery latency is very high. Furthermore, to overcome the problem of unnecessary message transmissions/receptions in epidemic routing, probabilistic flooding techniques(e.g., see ~\cite{BroadcastStorm,BulkDataDissemination} and references therein) have been developed for static ad hoc networks. Specifically, in~\cite{BroadcastStorm,BulkDataDissemination}, upon receiving a new message for the first time, each node (receiver) rebroadcasts the message once with a certain probability, which results in a significant reduction in the number of receptions compared to epidemic forwarding.
\tcb{Moreover, although a fishbone structure-based data dissemination method was proposed to analyze a forwarding path in a static and small-scale sensor network environment, this method still poses many challenging problems, including relay selection, when applied to a large-scale network environment~\cite{fishbone_wsn}.}

{\bf Data dissemination in mobile networks.} Several message forwarding algorithms have been presented for {\em mobile} delay-tolerant networks (DTNs)~\cite{CBR,SEBAR,Coff,SeeR}. The algorithms designed for mobile DTNs were built mostly based on contact-based routing strategies that exploit social behaviors while handling long delays and a dynamic network topology. 

{\bf Discussion.} While epidemic forwarding is known to perform quite satisfactorily in terms of coverage probability, its applicability to low-power IoT devices is questionable because a large amount of transmit power is caused by excessive broadcasting, yielding many duplicate message receptions at each node. Even if BIP was designed to reduce unnecessary message transmissions/receptions, the judicious design of an effective forwarding algorithm that maximizes the forward efficiency with an acceptable computational complexity so that it is suitable for ad hoc networks in which {\em low-power IoT devices} are deployed would remain an open problem. Because contact-based routing schemes in ~\cite{CBR,SEBAR,Coff,SeeR} utilize contact opportunities among mobile devices that are moving around in a rather small area, they are not applicable to our {\em large-scale static} network setting.

\section{Concluding Remarks} 
\label{sec_conclusion}

\tcb{In this study, we explored an open yet important problem of how to develop an effective data dissemination algorithm in the sense of both increasing the coverage probability and reducing the number of transmissions for massive IoT networks.} To address this challenge, we presented a new performance metric termed the forwarding efficiency and proposed the \textsf{FiFo} method, consisting of three phases, aimed at improving the performance metric with an acceptable computational complexity. More specifically, in Phase 1, we performed device clustering based on UPGMA; in Phases 2 and 3, we discovered a fishbone by creating the main axis and sub axes, respectively. Using a Twitter dataset, including our network topology, we demonstrated that our \textsf{FiFo} method is superior to the four benchmark data dissemination algorithms and exhibits an up to 74.54\% higher forwarding efficiency than the second-best performer (i.e., the modified BIP). Furthermore, we extended the proposed \textsf{FiFo} method to a case in which an additional relay was deployed over the network to avoid any potential coverage holes. In this case, we also validated the effects and benefits of our method with respect to coverage probability as well as forwarding efficiency.

Potential avenues for future research include the design of a more sophisticated data dissemination algorithm based on deep-learning models that exploit the topological information of the underlying network.



\ifCLASSOPTIONcaptionsoff
  \newpage
\fi

\bibliographystyle{IEEEtran}
\bibliography{IEEEabrv,references}

\end{document}